\documentclass[12pt]{iopart}
\usepackage{iopams}
\usepackage{graphicx}

\begin{document}
\topical{Spectroscopy and dynamics in helium nanodroplets}
\author{Frank Stienkemeier$^1$ and Kevin K Lehmann$^2$}
\address{$^1$Universit\"{a}t Freiburg, Physikalisches Institut, D-79104 Freiburg, Germany}
\address{$^2$Department of Chemistry, University of Virginia, Charlottesville VA 22904, USA}
\ead{frank.stienkemeier@physik.uni-freiburg.de}

\begin{abstract}
This article provides a review of recent work in the field of helium nanodroplet
spectroscopy with emphasis on the dynamical aspects of the interactions between molecules
in helium as well as their interaction with this unique quantum solvent. Emphasis is
placed on experimental methods and studies introducing recent new approaches, in
particular including time-resolved techniques. Corresponding theoretical results on the
energetics and dynamics of helium droplets are also discussed.
\end{abstract}

\submitto{J. Phys. B}
\section{Introduction}
Understanding the properties of matter starting from the interplay of atoms and molecules
has been achieved to a great deal by the study of small or model systems containing only
a few atoms. A detailed view on geometric as well as electronic properties has been
brought forward largely through application of spectroscopic tools. These tools are
continuously being improved, in particular with the aid of the availability of
sophisticated new laser systems, setting new milestones in terms of repetition rate,
power and the time and frequency structure of ultrashort pulses. On the other hand, the
path from small model systems to complex functional structures is arduous to climb. In
recent years, it has became clear that one approach towards the understanding of complex
structures of atoms and molecules is to start with well-defined structures and
well-defined distributions of populated states. This calls for spectroscopic studies at
very low temperatures and now a new field emerged dealing with cold molecules or
ultra-cold chemistry. In this regard, spectroscopic experiments involving helium droplet
beams (HElium NanoDroplet Isolation, HENDI) proved, since their introduction in 1992
\cite{Goyal:1992b}, to be a versatile method that provides temperatures below 1\,K and
offers the possibility to study well-defined and complex structures of atoms and
molecules. Moreover, the quantum nature and in particular the superfluid properties of
the droplets allow one to investigate these quantum properties in a size-limited
aggregate on the nanometer scale. Today the field of helium droplets is well established
and several review articles have appeared during the last decade highlighting a variety
of experimental as well as theoretical aspects
\cite{Whaley:1994,Vilesov:1998b,Whaley:2000,Northby:2001,
Callegari:2001,Sti:2001e,Ceperley:2001,Dalfovo:2001,Toennies:2001,
Krotscheck:2001,Toennies:2004,Jortner:2005,Barranco:2006}. In this paper, we will not
attempt a comprehensive overview of the entire field because many experimental as well as
theoretical approaches have already been reviewed, summarized and pointed out. Besides,
the field has grown to the point that a truly comprehensive review would now need to be a
monograph instead of a review article. We focus particularly on dynamical aspects of the
droplets and recent advances triggered e.g. by up-to-date femtosecond techniques on the
experimental side and on the theoretical side by significant progress in the
understanding of the dynamics and energetic properties.  We emphasize material that was
not available when the several reviews were written that appeared in the Special Issue of
the Journal of Chemical Physics
\cite{Northby:2001,Callegari:2001,Sti:2001e,Ceperley:2001,Dalfovo:2001} in late 2001.

The paper is organized as follows: We start with experimental aspects in the first part.
In particular, we summarize the conditions and parameters relevant for the formation and
doping of $^4$He as well as $^3$He droplet beams having certain size distribution. We
expect this information to be useful for a broad readership to get a feeling for
apparatus designs as well as feasibility considerations. In the next part, the properties
of helium droplets are reviewed, in particular focussing on energy, momenta and
excitation considerations which mainly determine the dynamics of e.g.~cooling and energy
dissipation processes. The last part reviews recent work in the field, particularly
addressing dynamical processes.

Much of the present knowledge about helium nanodroplets results from the work of
Roger Miller and his students at University of North Carolina, Chapel Hill.
We deeply regret the untimely passing away of Roger Miller.
Without his ideas and scientific activities, the field of helium droplets would be far
less advanced than it presently is. He was a colleague of unusual warmth, openness, and
enthusiasm, and arguably the most accomplished spectroscopist of his generation. In order
to admire his work we dedicate this Review to him.

\section{Experimental aspects}
\subsection{Production of helium nanodroplets and source design}
The formation of helium droplets in the size range starting from a few helium atoms up to
aggregates having centimeters in diameter has been achieved by a variety of techniques.
This review focuses on  nanodroplets in the size range of a couple of hundreds to ca.~a
million helium atoms which are produced as a beam in a supersonic expansion. In the
following, other approaches are shortly summarized.

Macroscopic drops having diameters from  1\,$\mu$m up to 2\,cm have been formed from
capillaries or as ``mist'' from the boiling liquid. These drops were levitated in
magnetic \cite{Weilert:1996} or laser dipolar \cite{Weilert:1995} traps. Because of the
much smaller surface to volume ratio compared to nanodroplets and thus limited rate of
evaporative cooling, these large drops have temperatures in the Kelvin range. Kim and
coworkers \cite{Kim:2000, Kim:2002} generated fog of superfluid helium by placing an
ultrasonic transducer beneath the surface of the helium liquid, generating drops on the
order of 100\,$\mu$m in size.  However, trapped large drops have not yet been utilized
for the isolation of molecules and spectroscopic studies. Ghazarian and coworkers formed
droplets in a pulsed helium expansion operated in the vapor above liquid helium
\cite{Ghazarian:2002}. Pick-up of neutral and charged copper atoms as well as electrons
was achieved from a laser plasma generated on a rotating disc at cryogenic temperatures.
In these experiments, the droplets are most likely not formed in the supersonic
expansion; the authors assign postexpansion condensation in the chamber filled with
helium at pressures on the order of 10\,mbar to be responsible for the agglomeration of
atoms to form droplets. Sizes are not given but likely to be larger compared to the
nanodroplets discussed in this review. Charged droplets having sizes in the micrometer
range have also been generated by electro-spray methods \cite{Tsao:1998}. Furthermore,
the possibility of extraction of charged droplets from a liquid surface was discovered
many years ago \cite{Boyle:1976}. Grisenti and Toennies have produced a jet of micron
size droplets by breakup of a superfluid liquid jet expanding into vacuum
\cite{Grisenti:2003}.

\begin{figure}
\resizebox{\columnwidth}{!} {\includegraphics{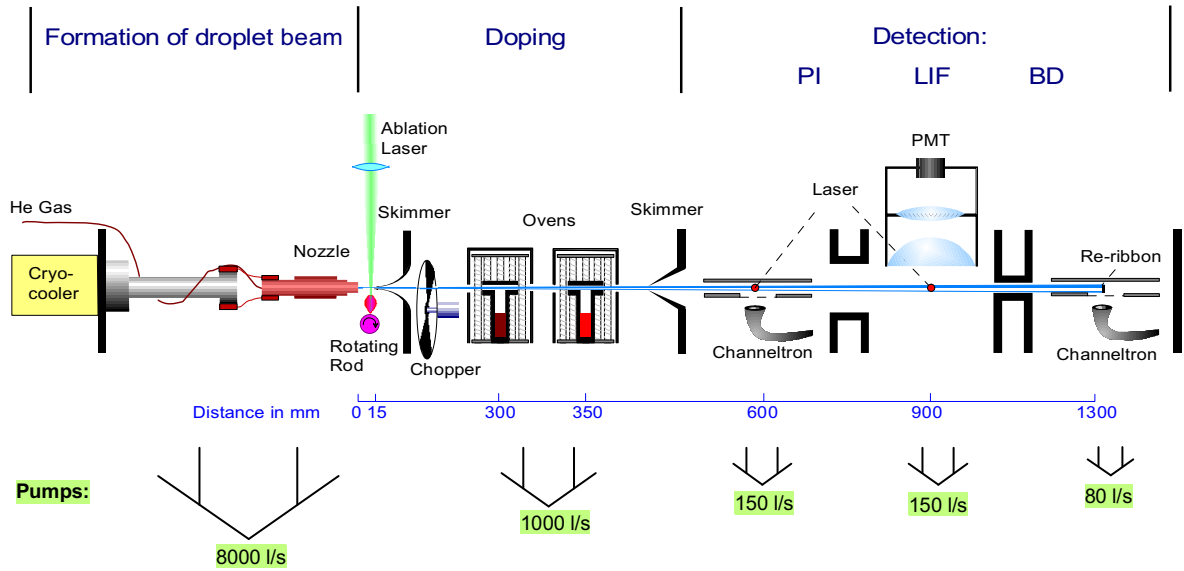}} \caption{Schematic view of a
helium droplet machine used for spectroscopic  studies. Depicted are the different vacuum
chambers for formation, doping and detection (PI: photo-ionization detection, LIF:
laser-induced fluorescence, BD: beam depletion by means of Langmuir-Taylor surface
ionization \cite{Sti:2000b}) of the droplet beam. Typical distances and required pumping
speeds are included.} \label{fig:apparatus}
\end{figure}

\begin{figure}
\center\resizebox{0.8\columnwidth}{!}{\includegraphics{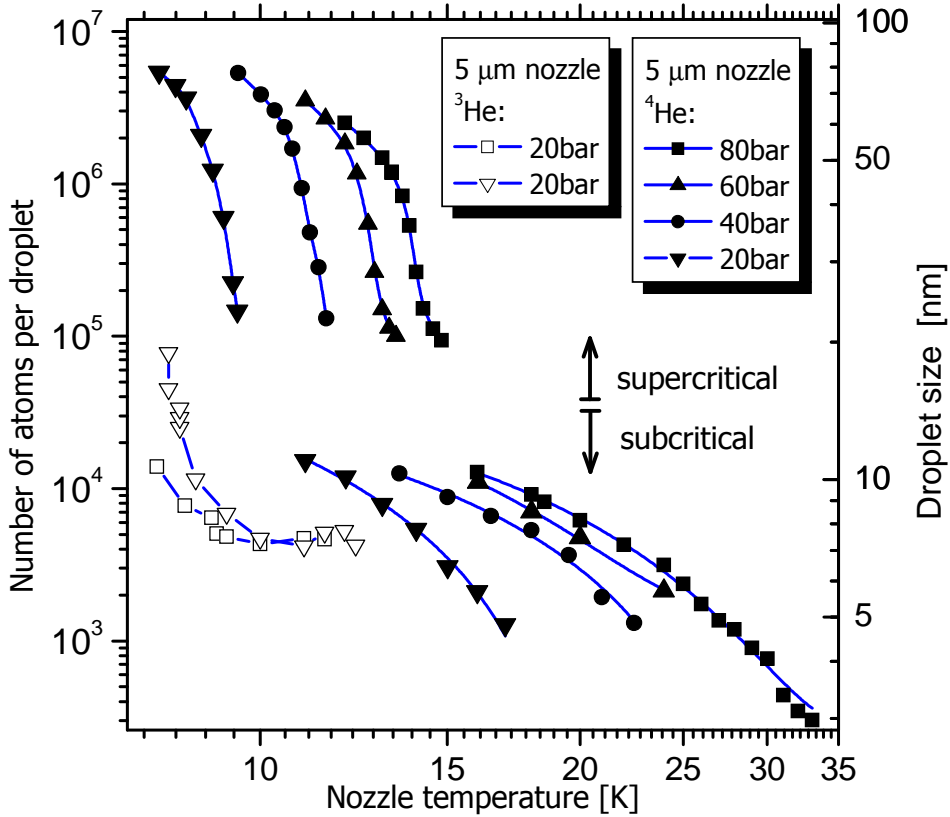}} \caption{Mean
droplet sizes $\bar{N}$ as a function of nozzle temperature T$_0$ for different
stagnation pressures P$_0$ measured by Toennies and coworkers using beam deflection. Data
points have been taken from \cite{Toennies:2004}. A nozzle having 5\,$\mu$m in diameter
has been used. The open symbols give sizes of $^3$He droplets determined by means of
scattering off a molecular beam \cite{Harms:2001} (triangles) or evaluating depletion
spectra \cite{Harms:1999a} (squares).} \label{fig:clustersizes}
\end{figure}

So far, all the experiments on the use of helium nanodroplets to spectroscopically study
attached species at millikelvin temperatures utilize supersonic expansion to form a
droplet beam freely travelling under high or ultra high vacuum conditions. One has to
establish nozzle temperatures T$_0$ below 40\,K and stagnation pressures P$_0$ in the
range of a few to 10\,MPa (100 bar) for the helium to condense to clusters or droplets.
Mean droplet sizes $\bar{N}$ of about 5\,000 He atoms per droplet, which is the standard
size for spectroscopic measurements, are formed under conditions of P$_0$ = 5\,MPa and
T$_0$ = 20\,K, when using a nozzle of 5\,$\mu$m in diameter (cf.
Fig.~\ref{fig:clustersizes}). With reasonably achievable pumping speeds in the source
vacuum chamber ($< 10\,000$ l/s) the nozzle pinhole openings have to be 20\,$\mu$m or
less to get a decent flux of droplets. Aperture discs for electron microscopes,
commercially available in high quality, are commonly used. A typical design of a helium
droplet machine is depicted in Fig.~\ref{fig:apparatus}. In most of the cases the source
chamber has an oil diffusion pump attached, providing high pumping speed. We use
unbaffled pumps 40\,cm in diameter specified with 8\,000\,l/s. However, in some cases we
also utilized turbo pumps providing 2\,000 -- 3\,000 l/s  with small nozzles. Droplet
fluxes were still on the same order of magnitude when compared to a setup having a large
diffusion pump.

\begin{figure}
\center\resizebox{0.8\columnwidth}{!}{\includegraphics{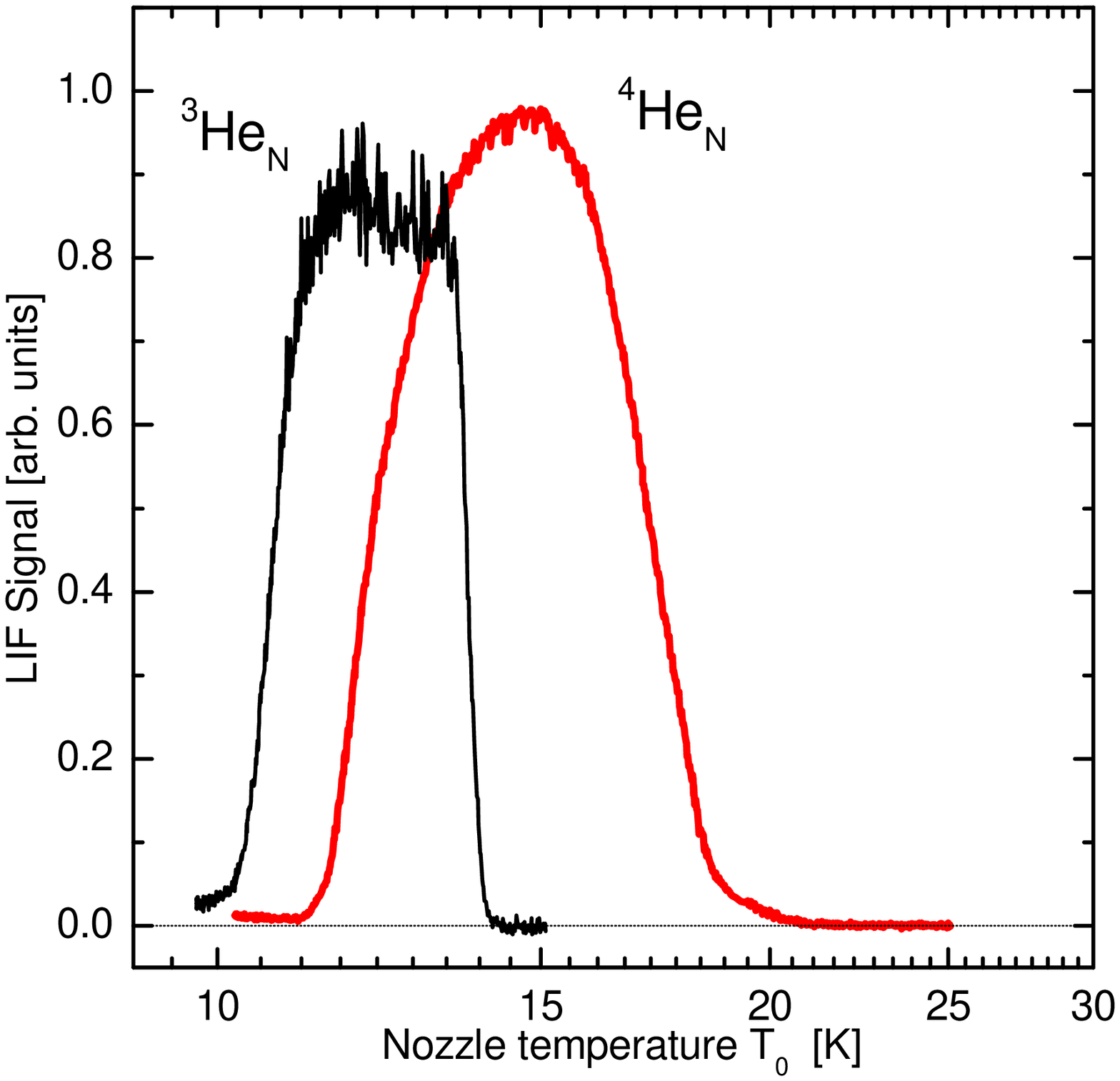}} \caption{Formation of
helium nanodroplets when cooling down the droplet source \cite{Sti:2004d}. The graph
compares absolute intensities of $^4$He (red) and $^3$He droplets (black), measured as a
fluorescence absorption signal of sodium-doped droplets.} \label{fig:cooldown}
\end{figure}

The process of helium droplet formation is well understood and discussed elsewhere
\cite{Toennies:2004,Toennies:1990,Harms:1996}. Starting at nozzle conditions P$_0$,
T$_0$, the droplets cool along an adiabatic line in the phase diagram (isentropic
expansion). The mean droplet size is determined by P$_0$, T$_0$ and the nozzle diameter.
Depending on whether one passes the critical point at the high or the low temperature
side, the droplets are formed by aggregating from the gas (subcritical expansion) or by
dispersing the liquid (supercritical expansion), respectively. These two regimes of
expansion parameters (nozzle conditions) are clearly identified in the resulting size
distributions (cf.~Fig.~\ref{fig:clustersizes}): supercritical expansion forms large
droplets ($\bar{N}\gtrsim 3\cdot 10^4$), whereas a subcritical expansion is suited to
generate droplets having $\bar{N}\lesssim 3\cdot 10^4$. Pushing nozzle temperatures below
$\approx$ 4\,K leads to very large droplets ($\bar{N} \approx 10^{10}$) with velocity as
low as 15\,m/s \cite{Grisenti:2003}. In this case Rayleigh instabilities break up the
liquid helium flow. The beam of droplets so produced had a divergence of only
$\approx$1\,mrad and thus had extremely high flux.

Absolute sizes have been measured by means of deflection in scattering experiments
\cite{Harms:1998b,Harms:2001,Lewerenz:1993} or attachment of electrons and deflection in
electric fields \cite{Knuth:1999}. The reason why most of the spectroscopic experiments
have been performed in the size range around 5000 helium atoms can be seen in
Fig.~\ref{fig:cooldown} where the droplets flux in terms of typical fluorescence signal
is plotted vs.~nozzle temperature. Decreasing the nozzle temperature increases the
droplet size. The maximum signal corresponds to the just mentioned size of about 5000 helium
atoms per droplet. Note the substantial lower achievable intensity in the supercritical
regime which starts for $^4$He droplets below 12\,K under the conditions used in
Fig.~\ref{fig:cooldown}.

As a rule of thumb, the divergency of the skimmed droplet beam is determined by the
geometry of skimmer opening and distance. We measured almost Gaussian profiles.
The beam is already several millimeters in diameter at a distance of 20\,cm  (e.g.
skimmer: 0.4\,mm in diameter, nozzle skimmer distance: 16\,mm).

All the characteristics and properties mentioned so far refer to the formation of
continuous droplet beams. A pulsed nozzle operated at high pressure and cryogenic
temperatures was successful used by the group of A.F.~Vilesov \cite{Slipchenko:2002}.
Compared to a continuous beam, orders of magnitudes higher droplet flux during the pulse
have been achieved. Pulse lengths have been determined to 30 -- 100\,$\mu$s. Apparently
this technique ideally combines with other inherent pulsed detection techniques, in
particular when short pulse lasers or time-of-flight techniques are involved. However,
due to shot-to-shot fluctuations, these pulsed beam sources so far are less attractive if
one is using a beam depletion detection method.

\subsection{Doping of droplets}
After having passed a skimmer, doping of the droplets is commonly achieved by inelastic
collision within a scattering cell, termed as the pick-up technique
\cite{Gough:1985,Scheidemann:1990}. The collision energies (velocities) involved in this
process are large compared to relevant energies of the superfluid (Landau critical
velocity).  Thus frictionless transport is not an issue and dopants are efficiently
trapped within the droplets. Pick-up cross sections have been determined to be on the
order of 50 -- 90\,\% of the total integral geometrical cross section of the droplets
\cite{Lewerenz:1995}. In terms of dissipation of translational energy, this means that
time scales for cooling have to be less than $\approx 100$\,ps \footnote{This time
corresponds to passing a droplet having 100\,{\AA} diameter at a velocity of 1000\,m/s.}.
Of course, to trap a solute, its translational energy must fall below the solvation
energy in that time, not reach equilibrium with the droplet. Typical lengths of the
scattering cells are a few centimeters. The required partial pressure of molecules in the
cell has to be on the order of about 10$^{-2}$\,Pa, providing a maximum of singly doped
droplets of $\bar{N} = 5000$. Depending on the material of interest, different techniques
are suitable to provide the required vapor pressure inside the scattering cell. Gases or
high vapor pressure liquids and solids samples are directly introduced through room
temperature capillaries. Higher temperatures are often used to establish the required
vapor pressure from bulk material inside a heated cell. Temperatures exceeding 1500\,K
have been used in order to evaporate metals \cite{Reho:2000a}. Thermal radiation from the
cell or heating are not a problem for the droplet beam because dipole active transitions
in helium require 20\,eV of photon energy. In the same way, high temperature assemblies
have been employed to dope radicals by means of pyrolysis \cite{Kupper:2002}. In this
case, an effusive continuous beam of radicals efficiently dopes the helium droplets.

One of us (FS) introduced laser evaporation as an alternative way to produce doped helium
droplets \cite{Claas:2003}. The material is ablated from a rotating and translating rod
by a pulsed YAG or Excimer laser. The laser plasma is produced several millimeters below
the beam near to the nozzle, which means doping takes place within the helium source
vacuum chamber. This location appears to be essential; apparently the high divergent flux
of helium atoms emerging from the nozzle opening is necessary to thermalize the large
translational energies of the atoms from the laser plasma. Metals like e.g.~Vanadium,
which are hard to evaporate thermally (required temperatures more than 1800\,K), have
been successfully loaded into the droplets in this way. The peak flux of singly doped
droplets proved to be as high as doping from an oven. Since high intensity nanosecond
laser pulses are required in order to produce the plasma, only a fraction of the
continuous droplet beam can be doped. Typically the doped fraction arrives at the
detector spanning about 100$\mu$s. Since a significant part of the laser plasma consists
of charged particles, also ions can be attached to helium droplets. For Mg one finds,
that the probability of finding positively charged droplets steeply increases with
increasing droplet size \cite{Claas:2003}. The time-of-flight mass distributions of these
ion-doped droplets revealed surprisingly large charged droplet distributions which
suggests that the presence of charged particles enhances the condensation of droplets.
The absolute fluxes of doped droplets passing a typical detection area of 1\,mm$^2$ at a
distance of 50\,cm is on the order of 10$^{10}$\,s$^{-1}$ corresponding to a number
density of 10$^8$\,cm$^{-3}$. Combination of laser evaporation with a pulsed helium
nozzle should substantially further increase the flux.

At elevated pick-up pressure, the droplets successively collect more than a single
atom or molecule. If one neglects the change in the droplet capture cross section with
number of atoms or molecules previously trapped, the probability of picking up $k$
species is expected to follow a Poisson distribution
$$P_k = \frac{\bar{k}^k}{k!}~e^{-\bar{k}}$$
where $k$ is the number of atoms or molecules picked up and $\bar{k}$ the mean number of
$k$, which is proportional to pickup cell pressure. This pressure dependence has been
verified for many molecules \cite{Lewerenz:1995,Nauta:2000b,Sti:2003}. If we neglect the
speed of the solutes relative to that of the droplet beam, we can approximate $\bar{k} =
\sigma \rho L$ where $\rho$ is the number density of the solutes in the pickup cell, $L$
the cell length and $\sigma$ the capture cross section (which is $\approx 0.15\cdot
N^{2/3}$\, nm$^2$ if one assumes the geometric cross section of the droplet).  Because of
the high mobility in the superfluid and the limited size of the confining droplet, the
individually collected atoms or molecules aggregate to form complexes or clusters
\cite{Lewerenz:1995}. When assuming $\sigma$ and $L$ to be constant these Poissonian
curves can directly be measured by varying the pick-up pressure. Such measurements are a
powerful tool to allow assignment of a spectroscopic feature to a certain oligomer sizes.
We applied this procedure successfully for clusters up to $k \approx 10$. Even the onset
of the intensity distribution allows an unambiguous assignment when recording intensities
of a complete series of cluster sizes $k$. The high pressure tails in the distributions
often significantly deviate from the Poisson functional form. In particular, the
assumption of a constant pick-up cross section $\sigma$ no longer holds at high pressure.
This is presumed to be due to the dissipation of the collisional, internal, and
complexation energies leads to evaporation of helium atoms from the droplet and a
subsequent shrinkage. In the case of dopants which are only weakly bound to the droplets,
desorption can drastically change the cluster size distributions as discussed
\textit{e.g.} in \cite{Vongehr:2003}.  Finally, the cross section $\sigma$ scales as
$N^{2/3}$. This means the optimal pickup density for doping a specific number of dopants
depends on the chosen mean He-droplet size. For different metals (alkali and alkaline
earth metals) the spectroscopic measurements done in our group (fs) and the group of
Meiwes-Broer in Rostock showed that one can agglomerate these atoms in a a helium droplet
without forming clusters unless they are laser-excited. A conclusive mechanism for such a
behavior is still lacking.

\begin{figure}
\center\resizebox{0.8\columnwidth}{!}{\includegraphics{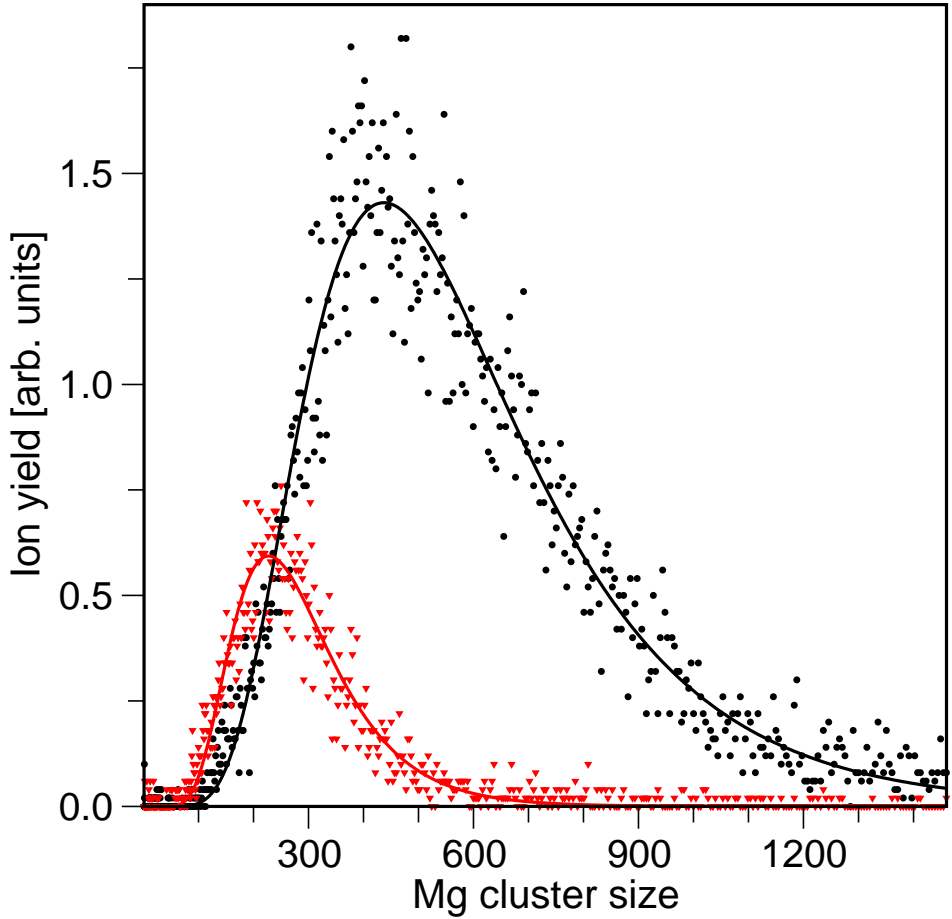}} \caption{Measured
magnesium cluster size distributions agglomerated in helium droplets. Data taken from
\cite{Diederich:2003}. Distributions are shown for two different densities of the Mg
vapor in the pickup cell, set by different cell temperatures of 520\,K (red triangles)
and 555\,K (black circles), respectively. The fitted log-normal functions (solid lines)
have maxima at cluster sizes of $k=227$ (437) and widths of 0.4 (0.46).}
\label{fig:oligomersizes}
\end{figure}

Due to the statistical character of the pickup process, experiments with a single cluster
size cannot be performed without further mass selection. The ensemble of droplets is
always loaded with a distribution of sizes. Since for small clusters the abundance of
individual cluster sizes can be extracted from the measured Poisson distributions, one
can experimentally verify cluster distributions also for larger clusters. As an example,
in Fig.~\ref{fig:oligomersizes} the size distribution of magnesium clusters is shown,
using photo-ionization and time-of-flight mass separation. Hence, if spectroscopic
features of different cluster sizes overlap, some ambiguity in the determination of
individual spectra cannot be avoided unless alternative means of cluster size separation,
such as dispersed emission, is used. Limitations in the number of picked-up constituents
are determined by the energy needed to dissipate collisional and binding energies on one
hand, and the energy ``capacity'' of the droplet on the other hand. Evaporation of one
helium atoms dissipates 5\,cm$^{-1}$ of energy \cite{Stringari:1987}. Droplets of size
5000 provide around 3\,eV energy for cooling capacity. Large droplets have been employed
to form metal clusters (e.g. Mg$_N$, Ag$_N$) containing several thousands of atoms
\cite{Diederich:2001,Doppner:2001,Diederich:2005}.

A unique strength in the formation of complexes and clusters in helium nanodroplets
lies in the ability to mix constituents arbitrarily and in a well defined order just by
placing pick-up assemblies in series along the droplet beam. In this way specifically
designed complexes containing a certain number of atoms and molecules can be synthesized
and studied. Just to name a few, studies have been performed on: Mg$_{1-3}$HCN
\cite{Nauta:2001g,Stiles:2003a}, Ne-, Kr-, Ar-HF \cite{Nauta:2001f}, Tetracene-Ar
\cite{Hartmann:1998}, HCN with  H$_2$, HD, and D$_2$ \cite{Moore:2001a,Moore:2003b},
Ag$_8$ with Ne$_N$,Ar$_N$,Kr$_N$,Xe$_N$ (N= 1-135) \cite{Diederich:2002}, NaCs and LiCs
\cite{Mudrich:2004}, HF-(H$_2$) \cite{Moore:2003a}, OCS-(H$_2$)$_N$ ($N=1-17$)
\cite{Grebenev:2000d,Grebenev:2001b,Grebenev:2002}.

In general the attached species are embedded inside the droplets forming compressed
helium solvation shells as discussed below. So far, only alkali atoms
\cite{Sti:1996a,Bruhl:2001b} and molecules \cite{Sti:1998a,Ancilotto:1995b}, and partly
alkaline earth atoms \cite{Sti:1997b,Sti:1999a,Reho:2000d,Mella:2005} are located on the
surface because of or their bubble-like solvation structures.
Ancilotto, Lerner, and Cole have presented a simple and successful model for
predicting the location of an atom or molecule bound to helium \cite{Ancilotto:1995e}.

\subsection{$^3$He droplet beams}
The study of $^3$He droplets adds in many experiments interesting aspects because of the
normalfluid property of these droplets compared to its superfluid $^4$He counterpart. The
lower chemical potential results in higher rates of desorbing atoms; thus the terminal
temperature of $^3$He nanodroplets in the evaporative cooling process amounts 150\,mK
\cite{Harms:1997b} which is considerably lower compared to 380\,mK for $^4$He droplets.
Since small $^3$He droplets appear not to be stable for sizes $N \lesssim 30$
\cite{Guardiola:2000a,Barranco:1997a,Guardiola:2000b}, aggregation is hindered because
droplets cannot be grown from small clusters adding one by one to an atom. In contrast,
cluster formation requires highly multiple collisions. As shown in
Fig.~\ref{fig:cooldown}, at comparable nozzle size and stagnation pressure, formation of
$^3$He droplets sets in at lower nozzle temperatures than for $^4$He droplets. Note
however, that the absolute intensities are quite comparable. Determined droplet sizes are
included in Fig.~\ref{fig:clustersizes} as open symbols. Surprisingly, there is some kind
of plateau region around $\bar{N} \approx 5000$; smaller sizes are missing. In the regime
of subcritical expansion, size distributions have been determined to be log-normal for
both $^4$He \cite{Harms:1998b} and $^3$He droplets \cite{Harms:2001}. Mixed $^3$He-$^4$He
droplets have also been formed by pickup of $^4$He by previously formed $^3$He droplets.
In these, the helium largely phase separates with the $^4$He located near the center of
the droplets.   The temperature of such mixed droplets is determined by the evaporative
cooling of the outer $^3$He layer.

\subsection{Spectroscopic tools}
Most spectroscopic tools and detection techniques that have been developed or used in
molecular beam studies are in principle applicable in combination with doped helium
nanodroplet beams. A large variety has already been applied and are described in previous
review articles \cite{Callegari:2001,Stienkemeier:2001,Toennies:2004}.   The most
commonly used have been Laser Induced Fluorescence (LIF) and depletion spectroscopy. LIF
is applicable and provides the greatest sensitivity, when atoms or molecules with high
quantum yield for optical emission are excited.  When using continuous wave laser sources
with sufficient intensity to approximately saturate the electronic transitions, photon
detection rates as high as $10^8$\,s$^{-1}$ have been observed in our experiments.  When
using pulsed lasers, the time averaged emission rate is reduced due to the small fraction
of the beam that can be pumped, $\approx 10^{-4}$ when using a 10\,Hz laser. Because of
the need for very high stray light rejection ($> 10^{12}$), careful baffling is needed
and multiple passing of the excitation light over the molecular beam is not practical.
One can use wavelength selection of the emission with a filter and (for pulsed
excitation) time gating to reduce the scattered light, which almost always dominates the
noise. Measurement of the wavelength dispersed LIF often provides insight into the
dynamics following absorption. Time correlated photon counting provides a way to study
sub-ns dynamics. Resonant multiphoton absorption and ionization has been used to great
effect as discussed below. Time resolved pump-probe spectroscopy using fsec pulsed lasers
have proved to a powerful tool for rapid dynamics.   While most studies have used laser
excitation sources, two groups have exploited synchrotron radiation in the deep UV, which
has allowed the study of the electronic excitation bands of the helium itself
\cite{vonHaeften:2001, vonHaeften:2002, vonHaeften:2005, Peterka:2003}.

When exciting ro-vibrational excitations in the IR or for many electronic excitations as
well, little or no fluorescence can be detected. In these cases, most of the energy
optically deposited in the droplet is ultimately lost by evaporation of helium atoms. As
mentioned earlier, a loss of about one helium atom per 5\,cm$^{-1}$ is expected based
upon the helium evaporation energy, though no quantitative experimental measurement of
this helium loss rate is yet available. The helium evaporation has been detected by
depletion of the on axis molecular beam, using either a bolometer or a mass spectrometer.
The bolometer provides greater sensitivity for beam flux detection.   The beam noise in a
one Hz bandwidth is equivalent to about one part in 50\,000 depletion of the beam, close
to the expected shot noise in the flux of droplets.  The excitation laser is typically
passed many times over the molecular beam using a pair of parallel, flat mirrors, so that
with kHz pulsed lasers a high duty cycle can be achieved.  Because of their low bandwidth
($\approx 250$\,Hz), bolometer detectors cannot be effectively used with gated detection,
for which mass spectrometer detection is advantageous. Depletion spectroscopy is more
demanding of power in the excitation source;  ideally one wants to use sufficient optical
intensity to have high probability of at least one excitation per doped droplet.

For vibrational transitions in the IR, the typical linewidth in helium is $\approx
0.5$\,GHz. If we assume a vibrational transition dipole moment of 0.1\,Debye and a
transition wavenumber of 3000\,cm$^{-1}$,  then the peak absorption cross section will be
$\approx 2 \cdot 10^{-16}$\,cm$^2$ and the excitation rate will be $\approx 4 \cdot 10^3
I$\,s$^{-1}$ where $I$ is the optical intensity in W/cm$^2$.  If 10\,cm of the beam, is
illuminated by the laser, this implies that the saturation intensity will be on the order
of 1 W/cm$^2$, and the saturation fluence is on the order of 250\,$\mu$J/cm$^2$.   The
situation in droplets is more complex, especially with cw excitation, due to the fact
that linewidth is often inhomogeneously broadened, and that there is spectral diffusion
between states and relaxation during the $\approx 250\,\mu$s that the molecules are being
pumped.

A particularly powerful spectroscopic tool, widely exploited by Roger Miller and
collaborators, has been the use of a strong DC electric field to strongly align molecules
in the laboratory frame while inside of helium droplets, producing what are known as
``pendular spectra" \cite{Nauta:1999c}.  In many cases, this leads to the collapse of the
rotational structure of a vibronic band into a single narrow Q branch like peak at the
band origin with an integrated intensity as high as three times that of the entire zero
field band.  This dramatically increases both sensitivity and resolution, particularly
important in the case where many closely related clusters are being formed in the
droplets \cite{Nauta:1999a, Moore:2001a}. Miller \textit{et al}. have also used this
technique to measure the angle between the static and transition dipole moments
\cite{Dong:2002}, which has proved to be a useful assignment tool for the spectra of
molecules with many conformers that each have multiple vibrational transitions in a
limited spectral region \cite{choi:2005a,choi:2005b}.  Strong alignment of electron
magnetic moments inside helium should also occur for modest magnetic fields ($\approx
1$\,T), and we anticipate this being exploited as more free radicals and their complexes
are studied in helium droplets.

\section{Energy, angular momenta and excitations in helium Droplets}
\subsection{Helium droplet excitations}
The excitations of helium nanodroplets play an important role in the thermodynamics and
much of the real time dynamics that occurs in pure and doped droplets.   As Landau
pointed out long ago \cite{Landau:1959}, the excitation spectrum of bulk helium is
responsible for its superfluid properties and it is natural to assume that the
excitations of droplets are key to understanding superfluidity (or even if that term has
any real meaning) on a nanometer scale.   The interaction of molecules with the helium
excitations allow molecules to be cooled inside the droplets and the rate of such cooling
is important in many chemical applications of HENDI.   Even after the molecule and
droplet reach equilibrium, the molecule-droplet excitation interactions determine time
scales for fluctuations and relaxations, and thus determine, among other things, the
homogeneous width of spectral lines.

Unfortunately, despite their importance, we have yet to develop experimental methods to
directly probe the excitation energies of helium nanodroplets, with the exception of
electronic excitations in the deep UV ($< 60$\,nm) \cite{vonHaeften:2001}.   In bulk
helium, the low energy excitations are mostly known from x-ray and neutron diffraction
experiments, but such measurements have not been done on droplets and are perhaps
unlikely given the relatively small helium droplet density presently available.    Thus,
much of what we know comes from theory, and in particular the liquid drop model
\cite{Brink:1990}, which allows one to make connections between the droplets and the well
known case of bulk helium \cite{Donnelly:1998}.

The liquid drop model \cite{Brink:1990} (LDM) treats a helium nanodroplet as a sphere
with the same density as bulk helium, $\rho = 21.8\,$nm$^{-3}$.  This implies a droplet
of $N$ helium atoms has a radius, $R(N) = 0.222\,N^{1/3}$\,nm.  It is known from Quantum
Monte-Carlo (QMC) calculations of various types \cite{Barnett:1993a, Cheng:1996} that the
ground and thermally averaged density profile of pure droplets is close to that assumed
in the LDM, except that the outer boundary is diffuse, with a 10-90\% width of $\approx
0.6$\,nm. The elementary excitations of such a sphere can be classified as two types,
ripplons and phonons. Ripplons are quantized capillary wave excitations.   They imply no
change in density of the droplet but a change of shape.   As such, their restoring force
constant is proportional to the surface tension, $\sigma$, which in the LDM is treated as
equal to the bulk value 0.363\,mJ/m$^2$.  These are characterized by an angular momentum
quantum number, $L \ge 2$.   A ripplon with angular momentum $L$ is a $2L +1$ degenerate
harmonic oscillator.   The excitation energy of this oscillator in the LDM is given by
$\sqrt{\frac{4 \pi \sigma L(L-1)(L+2) }{m \rho R(N)^3}  } = 3.77\,N^{-1/2}\,
\sqrt{L(L-1)(L+2)}$\,K. For large $L$, this goes over to the $k^{2/3}$ excitation curve
known for capillary waves on an flat surface \cite{Lamb:1916}. While ripplons have no
density change, except for the surface, there is a hydrodynamic flow field that scales as
$(r/R(N))^{ L}$ and thus becomes increasingly surface localized for higher $L$
\cite{Lamb:1916}. Ripplons are the lowest energy excitations of a nanodroplet and
dominate its thermodynamics below 1\, K \cite{Brink:1990, Lehmann:2003a}.

The other elementary excitations are phonons.   In the LDM, these are the compressional
normal modes of a sphere and are characterized by a radial quantum number, $n$, and an
angular momentum quantum number, $L$.  Phonons have amplitudes described by a spherical
Bessel function, $j_L(k_{nL} r)$; the values of $k_{nL} = r_{n,L} / R(N)$ where $r_{n,L}$
is the $n$'th root of $j_L$ \cite{Lamb:1916} .  In the LDM, the excitation energy of a
phonon with wavenumber  $k$ is assumed to be the same as that of a bulk helium excitation
of the same wavenumber; the complete empirical curve was given by Donnelly and Barenghi
\cite{Donnelly:1998}.   This excitation curve has a maximum of $13.76$\,K at $k =
11\,$nm$^{-1}$ (the maxon) followed by a minimum at $\Delta = 8.87\,$K at $k =
18.5\,$nm$^{-1}$.  Phonons with wavenumber near the minimum of the curve are called
rotons. The nature of these excitations have been discussed in \cite{Reatto:1999}. For $k
> 21\,$nm$^{-1}$, the excitation energy rapidly approaches $2\Delta$ due to hybridization
with two phonon states.   For $k$ significantly below the maxon, the excitation curve is
approximately linear with a slope given by the speed of sound in helium, $u = 236$\,m/s.
The lowest energy phonon is the symmetric breathing mode (n = 1, L = 0) which has an
excitation energy $25.5\,N^{-1/3}\,$K \cite{Brink:1990}.   Given the terminal temperature
reached by evaporative cooling of droplets \cite{Hartmann:1995a}, even the lowest phonon
is largely unpopulated for droplets with $N < 10^5$.   Ripplons dominate the nanodroplet
thermodynamic properties for temperatures below 1\,K, but the larger number of phonon
excitation states cause them to dominate the thermodynamics at higher temperatures.   At
such temperatures, however, the interactions between excitations are known to become
significant in the bulk.   It is likely that the phonon contribution to the thermodynamic
functions of the droplets at such high temperatures should be calculated from the
empirical bulk values \cite{Donnelly:1998}.   In doped helium droplets, it is known that
the solute will strongly modulate the helium density in the first solvation shell and
often in the second \cite{Dalfovo:1994,Kwon:1996}.  In addition, the spherical symmetry
will be lost for molecular solutes or even atomic solutes not in S atomic states.   The
phonon spectrum in the presence of solutes, even spherical, is largely unknown, but peaks
observed in the phonon wing of many electronic spectra have often been assigned to
phonons localized in the solvation structure around a solute \cite{Lindinger:2001a}, and
thus expected to be weakly dependent upon droplet size.   The energy of the lowest phonon
of each symmetry has been estimated using the POISTE method \cite{Huang:2004}, which
monitors the rate of decay of carefully selected trial functions under imaginary time
propagation \cite{Blume:1999}.   Energies are extracted using a maximum entropy
stabilized inverse Laplace transform.  A more general method, that calculates the normal
modes using time dependent, density functional theory is under development
\cite{Schmied:2005}  and will give the entire phonon spectrum, but with the limitations
of present density functionals which are unreliable for external potentials with very
attractive interactions. Krotscheck and coworkers have used variational Monte-Carlo
methods \cite{Chin:1995a} and quantum many-body methods known as HNC/EC
\cite{Chin:1995b}, Random Phase Approximation and Correlated Basis Function methods
\cite{Krotscheck:2001} to study the lowest excitations of each angular momentum in pure
and doped droplets.

Another type of `excitation' that should be considered is quantum evaporation
\cite{Wyatt:1992}.  In bulk helium, it is known that when a phonon/roton with excitation
energy greater than the atom binding energy ($7.2\,$K) strikes a vacuum interface,
coherent emission of an atom occurs. There is also a low energy phonon produced, with the
angles of the atom and phonon such that the momentum parallel to the surface is conserved
in the process.   In the LDM, the helium binding energy is reduced by a $11.23\,N^{-1/3}$
K due to a surface energy correction \cite{Hartmann:1999}.   The conservation of parallel
momentum is replaced by conservation of angular momentum, which will produce a
centrifugal barrier for the departing atom which will raise the threshold by  $\approx
1.23\,L(L+1)\,N^{-2/3}$\,K \cite{Lehmann:2004b}.  The time scale for quantum evaporation
can be estimated by the time it takes a phonon to transverse from the center to edge of a
droplet ($\approx 10$\,ps using the speed of sound) and this will lead to a lifetime
broadening of the phonons of $\approx 1$\,K.  This will largely ``wash out'' the discrete
excitation spectrum of the droplets and thus provide a true continuum into which
relaxation can take place. Given the low phonon density of states in nanodroplets, some
broadening mechanism is needed to rationalize the relatively smooth, narrow line shapes
of ro-vibrational transitions of molecules solvated in helium.

A vortex or vortex ring is a qualitatively different type of excitation that is possible
in nanodroplets \cite{Bauer:1979, Pi:2000}, but as of yet there is no evidence for their
presence \cite{Lehmann:2003b}.   A vortex is a topological defect with a quantum of
helium circulation around the defect.  The simplest such excitation is a straight line
vortex going down the middle of a droplet (north pole to south pole).  Such a vortex has
one unit of angular momentum per helium atom, and each atom a distance of $r$ from the
vortex has a velocity of $\hbar/ mr$.  In bulk helium, a hollow core model is quite
successful at predicting the energetics and velocity of quantized vortices
\cite{Rayfield:1964}. Applied to helium droplets, such a model predicts an energy of a
vortex in a helium nanodroplet of a few hundred K for droplets of interest in
experiments.   Extensive Density Functional Theory (DFT) calculations have been made of
the straight vortex in pure and doped droplets and the energetics are close to those
predicted by the simple hollow core model \cite{Pi:2000}.  The vortex angular momentum is
predicted to distort the droplet from spherical symmetry, but only slightly
\cite{Pi:2000,Lehmann:2003b}.  This energy of a vortex is so much higher than the droplet
temperature, that it was suggested that vortices would rapidly decay in nanodroplets.
However, such droplets are by far the lowest energy states of high angular momentum and
there do not appear to be any open channels by which such vortices could decay,
conserving both energy and total angular momentum, which is of course required
\cite{Lehmann:2003b}.   Furthermore, there are a whole family of curved vortex solutions
that have a vortex rotating around the droplet \cite{Bauer:1979}.  Both the energy and
angular momentum of such solutions drop as the distance of closest approach of the vortex
from the axis is increased.    To date, only hollow core model calculations have been
done on such solutions due to their lower symmetry \cite{Bauer:1979,Lehmann:2003b}.   It
is predicted that vortex solutions will be the lowest energy states of a droplet for
total angular momentum of 131 (846) $\hbar$ for droplets with $N = 10^3 (10^4)$
\cite{Lehmann:2003b}.  These values of the total angular momentum are quite a bit higher
than the RMS value expected for a droplet treated at a canonical ensemble at $T =
0.38\,$K (8.5 (38.7) $\hbar$ for $N = 10^3 (10^4))$.   However, recent statistical
evaporative cooling calculations that include angular momentum conservation have found
that a substantial fraction of the initial angular momentum deposited upon pick-up of a
solute atom or molecule (typically several thousand $\hbar$) is trapped in the droplets
after the helium evaporation is finished.  Thus, the doped droplets studied in
experiments should have sufficient angular momentum to produce vortex solutions.   In a
doped droplet with a vortex, the impurity is expected to bind to the vortex
\cite{Dalfovo:2000} (as is known to occur in the bulk) and could be expected to align
strongly with the vortex axis (though no quantitative calculation has yet been made).
Such a pinned solute molecule would not be expected to display the gas phase like
ro-vibrational structure that is almost always observed in helium droplets. It has been
predicted that the presence of a vortex will cause Ca atoms to move from surface to bulk
binding to a helium droplet \cite{Ancilotto:2003}.  The lack of observation of droplets
with trapped vortices is one of the most significant unresolved puzzles in the field.

For doped helium nanodroplets, another form of excitation is the translational motion of
the solute.   The minimum energy position of a solvated atom or molecule (with the
exception of alkalis and some alkaline earths atoms and their molecules) is in the center
of a droplet \cite{Ancilotto:1995e,Lehmann:1999}. If one assumes that the helium
solvation structure moves with the solute, then one can calculate the effective
confinement potential of the solute in terms of the long range He-solute C$_n$
constants~\cite{Lehmann:1999} and a buoyancy
correction~\cite{Lehmann:2000} due to the number of helium atoms displaced by the solute.
This leads to a 3D, approximately harmonic trap with vibrational frequencies typically
below 1\,GHz.  The translation motion of the solute results in a dense set of many
populated ``particle in a box'' states~\cite{Toennies:1995, Lehmann:1999}.   The direct
excitation of these modes upon optical absorption is not expected since the
predicted Doppler broadening is much  narrower than the frequency of the oscillator,
leading to strong motional narrowing~\cite{Lehmann:1999}.  Stated in another way, since
the absorbers are confined in a droplet much smaller than the excitation wavelength,
there expected Doppler broadening is Dicke-narrowed \cite{Dicke:1953} away. However,
there are at least three mechanisms that can lead to changes in the solute translational
energy upon optical excitation. One is that the potential of interaction of the solute
with the helium solvent changes upon electronic and/or vibrational excitation
\cite{Lehmann:1999}. Change in the helium solvation structure leads to a ``phonon wing'' in
the spectrum that is typically observed upon electronic excitation
\cite{Stienkemeier:1995c,Hartmann:1996a}. A change in the long range interaction
coefficients (particularly the C$_6$) will result in a difference in the effective
trapping potential between the two solute states in the transition.   This will result in
nonzero Frank-Condon factors for transitions that change the translational state and thus
fine structure on the absorption lines.   Another mechanism to ``light up'' the
translational motion states are coupling to the rotation of the
molecule~\cite{Lehmann:1999}.  A molecule at the very center of a droplet will experience
an isotropic potential. However, when it is displaced from the center, the dependence of
the C$_6$ on orientation of the helium relative to the molecule axis results in an
anisotropic potential that increases in magnitude rapidly as the solute is displaced
further from the center. There is another coupling that arises from the fact that the
effective translational mass of the solute is a function of its orientation and that this
leads to a hydrodynamic coupling of the orientation of the molecule with its
instantaneous momentum~\cite{Lehmann:1999}.   Both mechanisms lead to a mixing of
different translational states with molecular rotor states (especially those that differ
only in the angular momentum projection quantum number and thus are degenerate without
considering this interaction).  This translational state mixing leads to splitting of
what would otherwise be expected to be single solute ro-vibrational transitions. Since
these translational energy changes depend strongly on the size of the droplet, and almost
all experiments done to date sample a broad droplet size distribution, the expected fine
structure on solute transitions will likely be washed out, contributing to the line
broadening of the transition.

\subsection{Evaporative cooling and the temperature of helium nanodroplets}
In most HENDI experiments, droplets are doped by the pick-up technique pioneered by
Scoles and coworkers~\cite{Gough:1985}. The pick-up process will deposit substantial
kinetic energy and angular momentum into the droplets, in both cases approximately
proportional to the mass of the picked up species.   In addition, any thermal internal
energy of the dopant, the solvation energy of the dopant in helium, and any binding
energy (in the case of cluster formation) will contribute to the deposited energy.   If
we assume the droplets statistically distribute this energy, the temperature of the
droplets will raise to several K, typically above the $\lambda$ transition of bulk
helium.   Following this temperature jump, the droplets will rapidly cool by helium
evaporation~\cite{Brink:1990, Lehmann:2004b}.

Treatment of the many body evaporation by real time quantum dynamics is clearly a very
difficult problem.   Statistical rate treatments of evaporation are certainly feasible
and one can hope reasonably accurate, once the droplets have cooled sufficiently that the
evaporation rate is slow compared to the rate for equilibration between the excitations
of the droplet.   Unfortunately, we have no estimate for what that rate is!   The
statistical rate equations only require knowledge of the density of states as a function
of energy and the threshold energy for evaporation~\cite{Baer:1996}.   Brink and
Stringari~\cite{Brink:1990} made such a calculation, using the LDM to calculate the
density of states.   They predicted that for the evaporation time appropriate for HENDI
experiments, the terminal temperature should be $\sim 0.3$\,K.   Later IR HENDI
experiments, starting with Hartmann \textit{et al.}~\cite{Hartmann:1995a},  demonstrated
that the rotational populations of solvated molecules are well described by Boltzmann
distributions with temperatures in the range $0.37 \pm 0.02$\,K.  This result was
interpreted as strongly supporting the statistical evaporation model and also the natural
interpretation that the rotational populations of the solvated molecule were in
equilibrium with the droplet excitations resulting in the equal temperature.

Based upon the surface energy contribution to the evaporation energy, one would expect
that smaller droplets would cool to a lower temperature.   This is not the case, however,
because, at a minimum, the spread of droplet internal energy at least matches the
evaporation energy of the last helium atom, and this corresponds to a larger spread of
droplet temperature in a smaller droplet, which of course has a lower heat capacity.
Attempts to experimentally test the predicted dependence of final temperature on mean
droplet size is frustrated by the fact that the individual ro-vibrational transitions
broaden and become strongly asymmetric in shape for smaller droplet sizes, which makes
the determination of the temperature from the spectrum less precise~\cite{REMiller_pc}.

Most of the literature on helium droplets have, sometime tacitly, assumed that the
droplets are described by a canonical distribution at a final temperature near $0.38$\,K.
We cite, as examples, the predictions of spectral structure by one of the present
authors~\cite{Lehmann:1999} and the wide use of Path Integral Monte Carlo
calculations~\cite{Kwon:1996, Whaley:1998} which provide an average over a canonical
distribution (i.e. Boltzmann weighting) of states. However, it is clear that individual
droplets, after evaporation has stopped or slowed to a negligible rate, have a constant
energy, $E$, not constant temperature, i.e. are more properly described as a
microcanonical ensemble. Of course, it is possible for individual droplets to have
definite energy and for the ensemble of droplets probed in an experiment to follow a
canonical distribution.   However, the evaporative cooling calculations predict that for
droplet in the size range studied in most experiments, the spread of final energies for a
given droplet size is determined by the evaporation energy, not the energy fluctuations
of a canonical ensemble.   Perhaps more importantly, an isolated droplet must also
conserve total angular momentum, $J$, and thus should be described by an ensemble that
has both $E$ and $J$ as good quantum numbers~\cite{Lehmann:2004b}.

The evaporative cooling calculations of Brink and Stringari ignored angular momentum
constraints.  It is straightforward to include these in the evaporative cooling using a
formalism analogous to phase space theory in the treatment of unimolecular
dissociation~\cite{Baer:1996}.  Such calculations require knowledge of the density of
states as a function of both $E$ and $J$.   This density of states has been calculated
and fit to simple analytic functions for both ripplons and phonons (in the limit that
they can be treated with a linear dispersion relation)~\cite{Lehmann:2003a}.  Perhaps not
surprisingly, the density of states as a function of $J$ for fixed $E$ turns out to be
very accurately described by a Boltzmann distribution for a spherical top, but with an
effective temperature that has nothing to do with the droplet temperature.   The droplet
temperature, defined in terms of the derivative of the log of the density of states with
respect to energy at fixed total $J$, is a strong function of $J$ for fixed total $E$.
Monte-Carlo Statistical evaporative cooling calculations have been performed, conserving
total angular momentum, and the result was some what surprising~\cite{Lehmann:2003a}. The
final states are spread over a broad range of $E$ and $J$, but are clustered in a narrow
band around a line of constant temperature in the $E, J$ plane.  This final temperature
is essentially identical to that predicted by the earlier Brink and Stringari
calculations.   However, the spread of $E$ and $J$ is more than an order of magnitude
higher than predicted for a canonical distribution at the same temperature.   For fixed
$E$, the fraction of the initial angular momentum that is trapped in the droplet
increases with increasing $J$.   Also, the initial alignment of the angular momentum
(which for pick-up is primarily perpendicular to the droplet velocity vector) is largely
conserved.   Statistical theory predicts that this alignment is partially transfered to
the angular momentum of a solvated rotor.   Experiments had demonstrated a lab frame
alignment of tetracene, in qualitative agreement with these
predictions~\cite{Portner:2003}.  The statistical theory predicts that the evaporated
helium atoms are anisotropic in the frame moving with the droplet and measurement of that
distribution could provide a quantitative test of the statistical evaporation model.

The recent evaporative cooling results imply that the assumption of a canonical
distribution of internal states of the droplet and solute is likely a very poor
approximation.   This calls into question the accuracy of using the Path Integral Monte
Carlo for modeling helium droplets.  It also predicts that the distribution of center of
mass states of the solute is likely much broader than perviously predicted.   This may
explain, at least in part, the failure of theory to predict the inhomogeneous line
broadening that is commonly observed in IR spectra of doped helium
droplets~\cite{Lehmann:1999}.  An interesting consequence is that the high angular
momentum introduces a bias into the populations of a rotor such that the temperature
extracted from a fit to those populations in shifted higher than the temperature of the
entire droplet-solute system, treated as a statistical ensemble~\cite{Lehmann:2004a}. In
order to account for the strongly nonthermal effects that arise from the trapped angular
momentum, a realistic knowledge of the terminal distribution is needed.   While
statistical models can predict this, at present we have no way to assess if these
predictions are at all reliable.   Getting the final  temperature correct only provides
support that statistical theory is correct for the evaporation of the last few helium
atoms, which we would have expected to begin with. Clearly, there is a important need for
experiments that can measure the distribution of the the droplet excitations.   At the
end of evaporative cooling, most of the trapped angular momentum is predicted to be in
the lowest ripplon mode, creating a quadrupole distortion of the droplets.   It is at
least in principle possible that excitation in this mode could be probed with Raman
Spectroscopy (though signal estimates are not encouraging!) or by atom scattering
experiments, particular by atoms held e.g.~in a magneto optical trap.

\section{Superfluidity and Molecular Rotation in Nanodroplets}
Superfluidity in helium has continued to fascinate since its discovery in the 1930's. The
phenomenological two fluid model is of great value in interpreting experiments
\cite{Tisza:1938}. In this model, helium below the $\lambda$ point is treated as a
mixture of normal and superfluid liquids.   Landau developed the theory of the normal
fluid component as a gas of quasiparticle excitations \cite{Landau:1959}. He also gave an
experimental definition of the normal fluid fraction in terms of the ratio of the
observed to classical moment of inertia for infinitesimal rotation of the
sample, a definition used by Andronikashvili in his classic experiment with a torsional
oscillator \cite{Andronikashvili:1946}. Ceperley developed a Path Integral Monte-Carlo
(PIMC) estimator that calculates this inertial response of a sample in a simulation
cell~\cite{Ceperley:1995}.  This involves calculating the projected areas of Feynman
loops that include He exchange. Superfluidity as defined by Landau is a macroscopic
concept; Feynman explicitly warns against giving it a microscopic
interpretation~\cite{Feynman:1990}.  Yet, the appeal to interpret behavior of helium
nanodroplets in terms of a superfluid has been irresistible.  Nanodroplets provide a way
to study superfluidity on a nanometer length scale, but as is common with other bulk
thermodynamic quantities, ambiguities arise when applied on the atomic scale.

Sindzingre, Klein and Ceperley \cite{Sindzingre:1989} provided the first calculations of
the normal and superfluid fractions of a helium cluster.   They found clusters as small
as 64 He displayed clear evidence for superfluidity based upon the inertial response, and
clusters of 128 atoms displayed a remnant of the $\lambda$ transition. Similar
calculations where then done on a 39 atom helium cluster doped with SF$_6$
\cite{Kwon:1996}. The observation that molecules often have substantially reduced
rotational constants when observed in helium nanodroplets was interpreted as a
``Microscopic Andronikashvili Experment'' \cite{Grebenev:1998}, with the increase in the
moment of inertia assigned to a helium normal fluid that rotates with the molecule.  In
order to calculate such a predicted moment of inertia, one needs the spatial dependence
of the normal fluid density, which is not provided by Ceperley's original method.  Kwon
and Whaley \cite{Kwon:1999b} introduced a PIMC estimator of a local normal fluid density
based upon a separation of the number of short and long exchange paths. Assuming that
this ``normal fluid'' rotated rigidly with the molecule (what they referred to as
adiabatic following), they calculated the helium contribution to the effective moment of
inertia of SF$_6$ in helium and found good agreement with the experimentally determined
value \cite{Kwon:1999b}. This group published several papers using this approach
\cite{Kwon:2000, Kwon:2001b, Kwon:2003}.

This definition of local normal fluid has been questioned by Draeger and Ceperley
\cite{Draeger:2003}.  The derived local density, when integrated, does not reproduce the
inertial properties calculated with the Ceperley PIMC estimator, i.e. it does not give
the proper global normal fluid fraction.  It is a scalar quantity, while the inertial
response and thus normal fluid fraction of an inhomogeneous quantum fluid is a second
rank tensor as is the Ceperley PIMC estimator.
Draeger and Ceperley introduced new definitions of local normal and superfluid densities
that have the proper symmetry and integrate to the correct global densities.  They used
the normal fluid estimator to calculate the moments of inertia for rotation of (HCN)$_n$
clusters in helium \cite{Draeger:2003} by assuming this normal fluid rotates with the
molecule.  One interesting conclusion of their work is that there is a normal fluid
fraction for rotation around the symmetry axis that reflects thermal excitations in the
first solvation layer present even at 0.38\,K. Such excitations carry angular momentum
and result in the prediction of a Q branch in the spectrum.   Such Q branches have been
seen in the spectrum observed by the Miller group \cite{Miller:2001} and their relative
intensities are in at least qualitative agreement with the predictions of the normal
fluid fractions calculated~\cite{Draeger:up}. The Draeger/Ceperley estimator is not
unique.   One could add to the definition a second rank tensor that will
integrate to zero and still recover the proper global superfluid estimator. Recently,
Kwon and Whaley \cite{Kwon:2005} have introduced such a modification and obtained local
superfluid densities quite different from the Draeger/Ceperley estimator.

Another approach to the calculation of the effective rotational constants of molecules in
helium is to calculate the hydrodynamic flow of helium needed to maintain a static helium
solvation structure in the rotating from of the molecule
\cite{Callegari:1999,Callegari:2000a}. This treatment assumes adiabatic following of the
helium density structure that rotates rigidly with the molecule (but not the helium atoms
themselves). This assumption, combined with the equation of continuity, gives a
differential equation for the helium velocity.  It is further assumed that the helium
flow is irrotational, which Lord Kelvin showed was the minimum kinetic energy solution to
the equation of continuity for a given time dependent density \cite{Lamb:1916}.  This
implies that the local helium velocity can be written as minus the gradient of a scaler
function known as the velocity potential. This definition of adiabatic following of the
quantum hydrodynamic model is distinct from that introduced independently by Kwon and
Whaley \cite{Kwon:1999b} and is based upon an assumption that the frequency for molecular
rotation is low compared to the excitations of the helium. The normal fluid response
calculations are based upon infinitesimal rotation and thus is also dependent upon an
assumption of a separation of time scale. The helium density structure was estimated
using helium Density Functional Theory \cite{Dalfovo:1995}, though this approximation was
not essential.   Good agreement was found for the rotational constants of a number of
heavy molecules (especially considering the considerable uncertainty of the He-molecule
interaction potentials), but not for light molecules HCN and HCCH.  The later were
attributed to a breakdown of adiabatic following, which was later demonstrated
experimentally \cite{Conjusteau:2000} and theoretically \cite{Viel:2001} in the case of
HCN.

A similar hydrodynamic calculation for the hydrodynamic mass of alkali cations in helium
was found to be in essentially quantitative agreement \cite{Lehmann:2001c} with high
level QMC calculations \cite{Rossi:2004}. Kwon \textit{et al.} \cite{Kwon:2000} reported
that the hydrodynamic model dramatically underestimates the effective moment of inertia
of SF$_6$ in helium. Lehmann and Callegari \cite{Lehmann:2002} agreed with this
conclusion if one used previously reported PIMC densities as input to the calculation,
but found that if Diffusion Monte-Carlo (DMC) densities (calculated with the same
potential!) were used, the hydrodynamic calculations were in quantitative agreement with
experiment. They argue that since the hydrodynamic model is a zero temperature theory,
the ground state helium density (which are calculated with DMC) should be used instead of
the thermally averaged density (which is what is calculated by PIMC).

For light rotors (B $\ge$ 0.5\,cm$^{-1}$), adiabatic following breaks down and one must
explicitly account for molecule rotation in any quantum simulation. Viel and Whaley
\cite{Viel:2001} used the POITSE method to estimate the energy of excited rotational
states of HCN. The Whaley group has continued the development of this method and has
applied it to a number of systems \cite{Paesani:2004b, Paesani:2005a, Paesani:2005b}.
More accurate is the use of imaginary time orientational correlations functions to
extract the excited rotational states which were introduced by Moroni \textit{et al.}
\cite{Moroni:2003, Moroni:2004, Cazzato:2004, Paolini:2005, Moroni:2005}. This has
allowed for essentially exact calculation of the rotational excitation energies of
molecules in small to modest size helium clusters and has been found to be in essentially
quantitative agreement with recent experiments (discussed below). Blinov and Roy
\cite{Blinov:2004, Blinov:2005} have introduced similar calculations using PIMC instead
of DMC to calculate the orientational correlation functions.   The Whaley group is also
using similar methods \cite{Zillich:2004b, Zillich:2005} in addition to several
others that they have pioneered.

Zillich and Whaley have developed the DMC/CBF method for treating the interaction of
molecular rotation excitation with the phonons and rotons of helium and applied it to the
case of HCN \cite{Zillich:2004a} and C$_2$H$_2$ \cite{Zillich:2004b}, an important yet
challenging problem. This model treats the phonons of the droplet as if they are
continuous and have the same form as for bulk helium and then decomposes the density
anisotropy that rotates with the molecule in terms of these phonon modes.
In this model, off resonant interactions shift the rotational excitation energy.
The interaction of the rotor with the rotons is predicted to lead to absorption intensity
for the rotons.  The observation of such absorption would provide an important
further test of the DMC/CBF model.

The use of Density Functional Theory for treating helium has been very active and is
beyond what we can realistically cover in this review.  The groups of F. Ancilotto, M.
Barranco, F. Dalfovo, J. Eloranta, E.S. Hernandez, and L. Szybiryz have been particularly
active. A great part of this work has been reviewed in \cite{Barranco:2006}.

\section{Review of recent work}
\subsection{Femtosecond dynamics}
Dynamical processes are directly accessible applying femtosecond real-time spectroscopy.
The first explicitly time resolved studies in helium droplets used the technique of time
correlated photon counting, which allowed detection of time resolved changes in emission
with $\approx 100$\,ps time resolution.  This proved sufficient to detect radiative and
quenching rates and the exciplex formation rates in cases of a sufficiently high barrier.
Ultrafast laser pump/probe experiments allow probing time scales from tens of
femtoseconds up to hundreds of picoseconds.  These methods allow the nuclear motion of
the probed atoms and molecules and changes in solvation structure to be addressed. In
connection with the experiments in helium droplets this is of peculiar interest because
of the quantum nature at these low temperature and effects related to e.g.~delocalization
of atoms or the superfluid properties, which outstandingly alter the dynamics of a fluid.
In collaboration with C.-P.~Schulz (Max-Born-Institute Berlin) one of our groups (FS) has
performed over several years femtosecond pump-probe experiments of doped helium
nanodroplets. It was demonstrated that even with the dilute target of a doped droplet
beam, dynamical studies can be successfully realized. A variety of problems have been
tackled which are summarized in the following.

\subsubsection{Wave packet propagation of alkali dimers}
\begin{figure}
\center\resizebox{0.8\columnwidth}{!}{\includegraphics{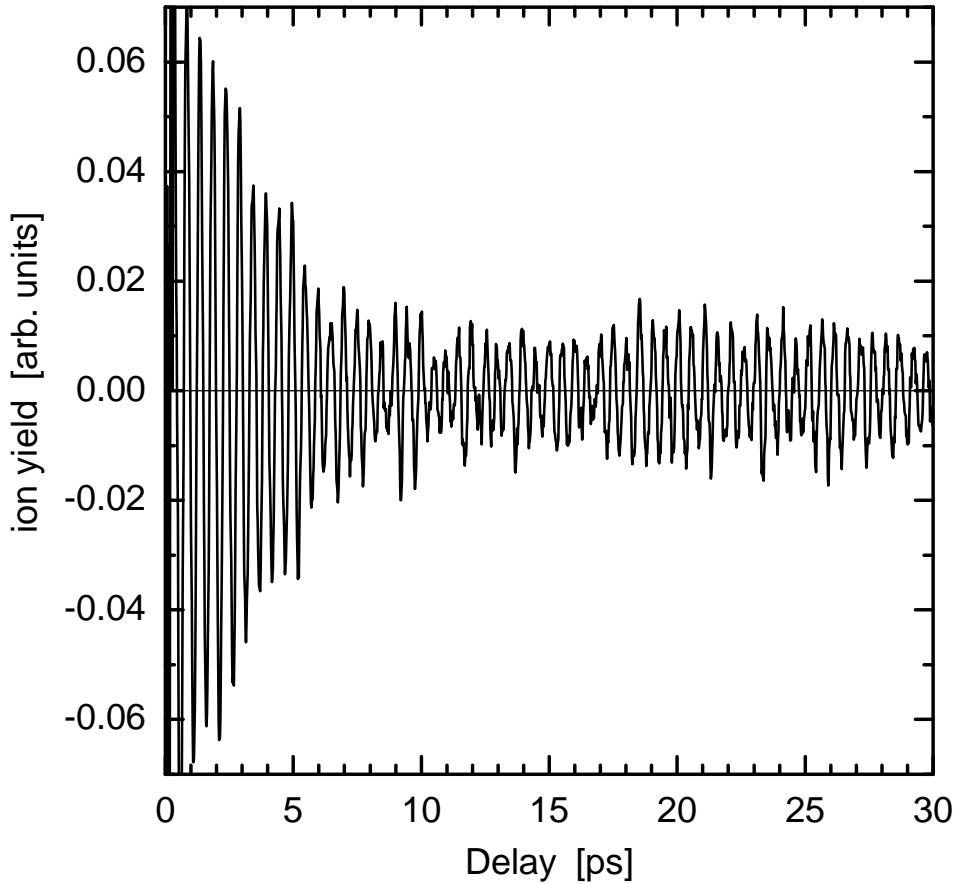}} \caption{Wave
packet propagation of potassium dimers attached to helium nanodroplets. Pump and probe
pulses at 12000\,cm$^{-1}$ were used. The offset present in the original measured trace
was subtracted to center the oscillation around the origin. The oscillation essentially
represents the vibrational motion in the $A^1\Sigma_u^+$ state. }
\label{fig:K2fsoscillation}
\end{figure}

\begin{figure}
\center\resizebox{0.8\columnwidth}{!}{\includegraphics{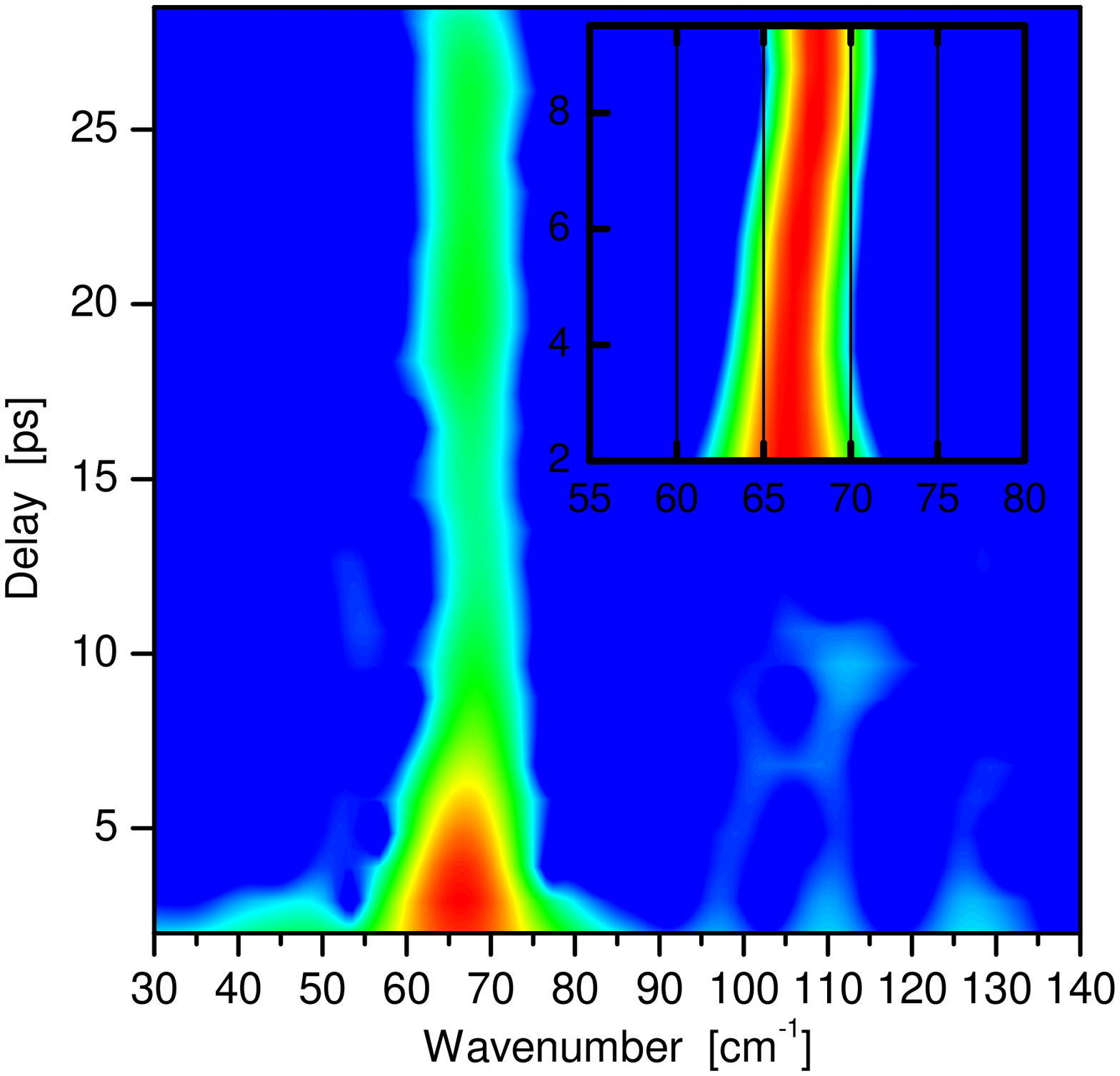}}
\caption{Spectrogram of the pump-probe oscillation shown in
Fig.~\ref{fig:K2fsoscillation}. The main intensity around 66\,cm$^{-1}$ corresponds to a
wave packet in the  $A^1\Sigma_u^+$ state of potassium dimers attached to helium
nanodroplets. The inset contains the same data, but normalized for each delay time in
order to illustrate the shift of the vibrations during desorption of the dimer from the
helium droplet. } \label{fig:K2fsvibshift}
\end{figure}

The goal of studying the geometric and electronic structure of molecules at low
temperature conditions relies on the weak and isotropic interaction with the helium
environment. Results  and their interpretation may significantly degrade if one is not
able to quantify the perturbation of the surrounding matrix. For that reason the
vibrational motion of molecules attached to helium droplet has to be compared to
gas-phase experiments. Vibrations of alkali dimers attached to helium nanodroplets were
followed in real-time applying femtosecond pump-probe techniques. Depending on parameters
like the excitation wavelength or photon intensity, wave packet propagation in potassium
dimers has been observed in the electronically excited states $A^1\Sigma_u^+$ and
$2^1\Pi_g$, as well as in the singlet ground state $X^1\Sigma_g^+$ \cite{Sti:unp_fsK2},
see for example Fig.~\ref{fig:K2fsoscillation}. The oscillation of the mass-selected
photo-ionization intensity directly images the wave packet motion revealing vibrational
frequencies as well as its time dependence. A Fourier analysis of the time spectra
comprises the assignment to the excited vibrational states. Furthermore, the phase of the
oscillations gives information on transition probabilities and the location of the probe
window, i.e.~the distance at the internuclear coordinate which amplifies the probe step
and is required for detecting dynamics. When compared to the free dimer, the formation of
a ground state ($X^1\Sigma_g^+$) wave packet from a Raman process was found to be
strongly enhanced in the droplet environment. At photon energies around 12500\,cm$^{-1}$
this process even dominates the pump-probe signal. Since the vibrational motion in the
ground state was found to be unaltered when compared to the gas-phase K$_2$, the
difference in the transition probabilities  must originate from the perturbed
electronically excited states. Indeed, the wave packet propagation in the electronically
excited $A^1\Sigma_u^+$ state clearly discovers a reduced vibrational wavenumber by
ca.~1\,cm$^{-1}$. The influence of the helium matrix leads to a wider vibrational level
spacing in that part of the excited state potential. Time resolved Fourier analysis
(spectrogram technique \cite{Schreiber:1997}) allows one to follow the vibrational
energies approaching gas-phase values during desorption of the excited potassium dimer
from the helium droplet.  In Fig.~\ref{fig:K2fsvibshift} a corresponding spectrogram is
presented. The inset clearly illustrates the shift to higher frequencies upon desorption.
Furthermore, the decrease in amplitude of the oscillation is noteworthy. It is either
associated with decoherence effects upon desorption or in changes of the Franck-Condon
detection window. In this regard desorption times upon electronic excitation have been
determined for the first time. The experiments give a desorption time of K$_2$ excited
into the electronic A state of $\approx 3$\,ps. The analysis of excited wave packets in
higher electronic states of K$_2$ reveals an even faster desorption behavior
\cite{Sti:unp_fsK2}.

Exploiting the advantage of a preferential formation of \emph{weakly} bound alkali dimers
and clusters \cite{Sti:2004}, for the first time wave packet propagation of triplet dimer
states has been observed. Experiments on sodium dimers attached to helium droplets
allowed the observation of vibrational wave packets in the $1^3\Sigma_u^+(a)$ as well as
$1^3\Sigma_g^+(c)$ states \cite{Sti:unp_fsNa2} which confirm former measurements
performed in the frequency domain and support the theoretical calculations on the dimer
interaction potentials.

\begin{figure}
\center\resizebox{0.8\columnwidth}{!}{\includegraphics{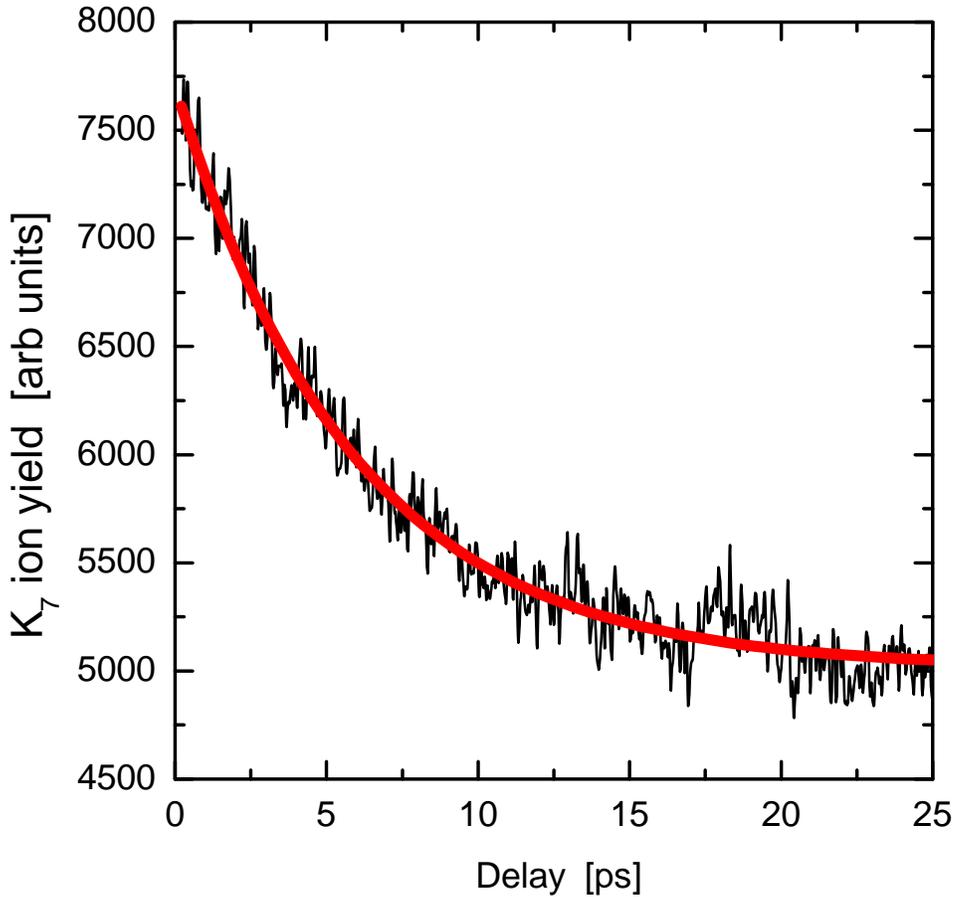}} \caption{Pump-probe
spectrum of potassium clusters attached to helium nanodroplets recorded on  the mass of
K$_7$. The fit according to a model taken from \cite{Kuehling:1993} deduces a decay time
of K$_7$ attached to helium droplets of 5.8\,ps. A photon energy of 12195\,cm$^{-1}$ was
used. } \label{fig:Knfsdecay}
\end{figure}

\begin{figure}
\center\resizebox{0.8\columnwidth}{!}{\includegraphics{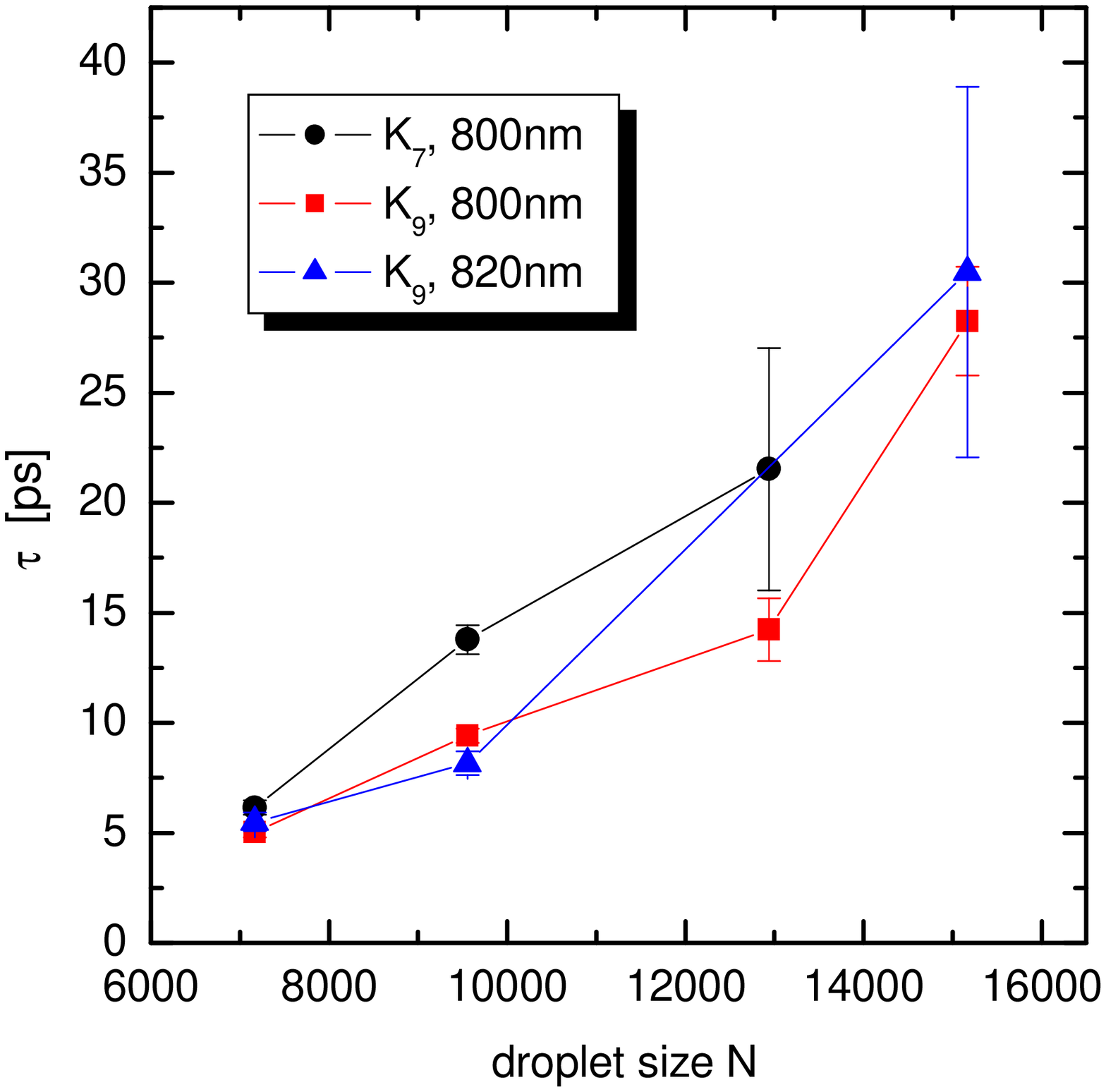}} \caption{Determined
fragmentation times as a function of the mean helium droplet size. The significant
droplet size effect indicates the much more efficient cooling mechanism of the larger
droplets. } \label{fig:Knfssizeeffects}
\end{figure}

\subsubsection{Fragmentation dynamics of alkali clusters}
The fragmentation dynamics of alkali clusters (K$_n$, $3<n<11$) has been studied in order
to directly measure the energy dissipation and cooling process of helium droplets
\cite{Sti:unp_fsKn}. These experiments were performed using small potassium clusters
because their electronic excitation and subsequent fragmentation has been investigated in
a variety of experimental studies. The pump-probe experiments of alkali-cluster doped
helium droplets allow one to follow the abundance of selected cluster sizes in real-time.
In this case the probe step determines the yield of mass selected ions from a
non-resonant photo-ionization process. Fig.~\ref{fig:Knfsdecay} plots such a recorded ion
yield as a function of delay time. Following a fragmentation model introduced by
K{\"u}hling et al.~\cite{Kuehling:1993}, a fragmentation time of 5.8\,ps is determined in
this case. In the gas-phase potassium as well as sodium show values which are around
$\approx 1$\,ps \cite{Kuehling:1994}. The longer fragmentation times determined in
connection with helium droplets are interpreted in terms of a competing channel where the
energy of the laser-heated alkali cluster is dissipated into the helium droplet prior to
fragmentation. Fig.~\ref{fig:Knfssizeeffects} shows a striking droplet size effect of the
measured fragmentation dynamics. The larger droplets cool the energy much more
efficiently, resulting in even longer fragmentation times. The energy to be dissipated
depends on the photon energy, cluster size etc.~and will be discussed thoroughly in
\cite{Sti:unp_fsKn}. Roughly speaking, excess energy of the order of 1\,eV is transferred
to the helium droplet in less than a ps to comply with the observed effects. Assuming
that this energy (thousands of K) cannot reside in internal degrees of freedom of the
droplet and is ad hoc evaporatively cooled, this means, that on the order of 2000 helium
atoms per ps leave the droplet. Experiments on NO$_2$ excited well above the gas phase
dissociation threshold but with no evidence of dissociation in helium also suggest
vibrational relaxation cooling with similar ultrafast rates \cite{Conjusteau_thesis,
Stolyarov:2004}.

\subsubsection{Surface dynamics}
In the same line of experiments, recent results demonstrate that alkali atoms can be
utilized as probes for surface excitations of helium droplets
\cite{Sti:1999b,Sti:unp_fsCs}. Inducing a spacial expansion of the alkali valence
electron orbital upon laser excitation, the helium environment rearranges. Several
time-dependent features are identified in pump-prob experiments. Assignment to specific
surface modes of the droplets could so far not be achieved. As expected, strong droplet
size effects are present. The expansion of the dimple structure takes place at picosecond
time scales. In these experiments normalfluid $^3$He droplets have been directly compared
to superfluid $^4$He counterparts. Surprisingly $^3$He droplets show a comparable
behavior and quite similar effects.

\subsubsection{Exciplex formation}
\begin{figure}
\center\resizebox{0.75\columnwidth}{!}{\includegraphics{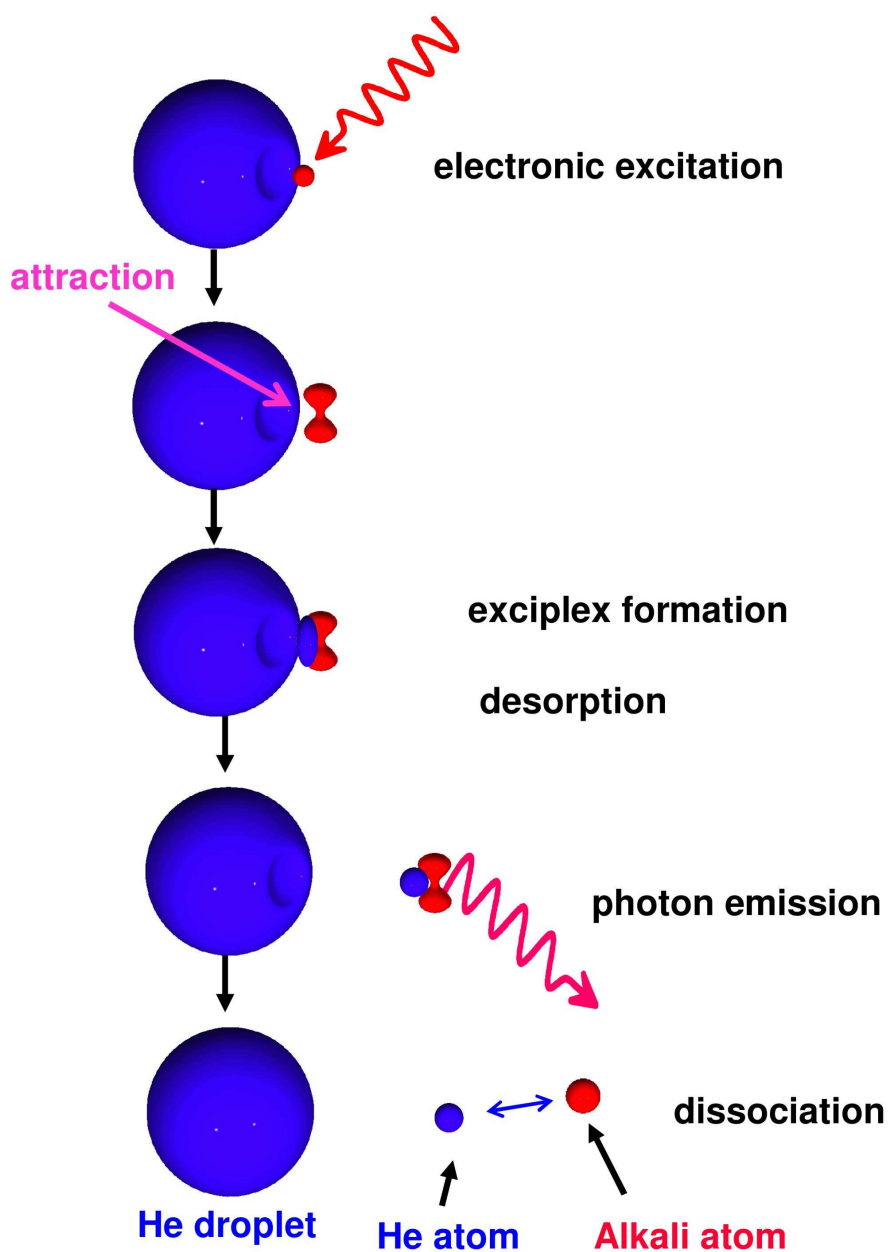}}
\caption{Excitation of an alkali atom attached to the surface of a helium nanodroplet and
subsequent formation of an excited alkali-helium molecule. } \label{fig:exciplexes}
\end{figure}

\begin{figure}
\center\resizebox{0.8\columnwidth}{!}{\includegraphics{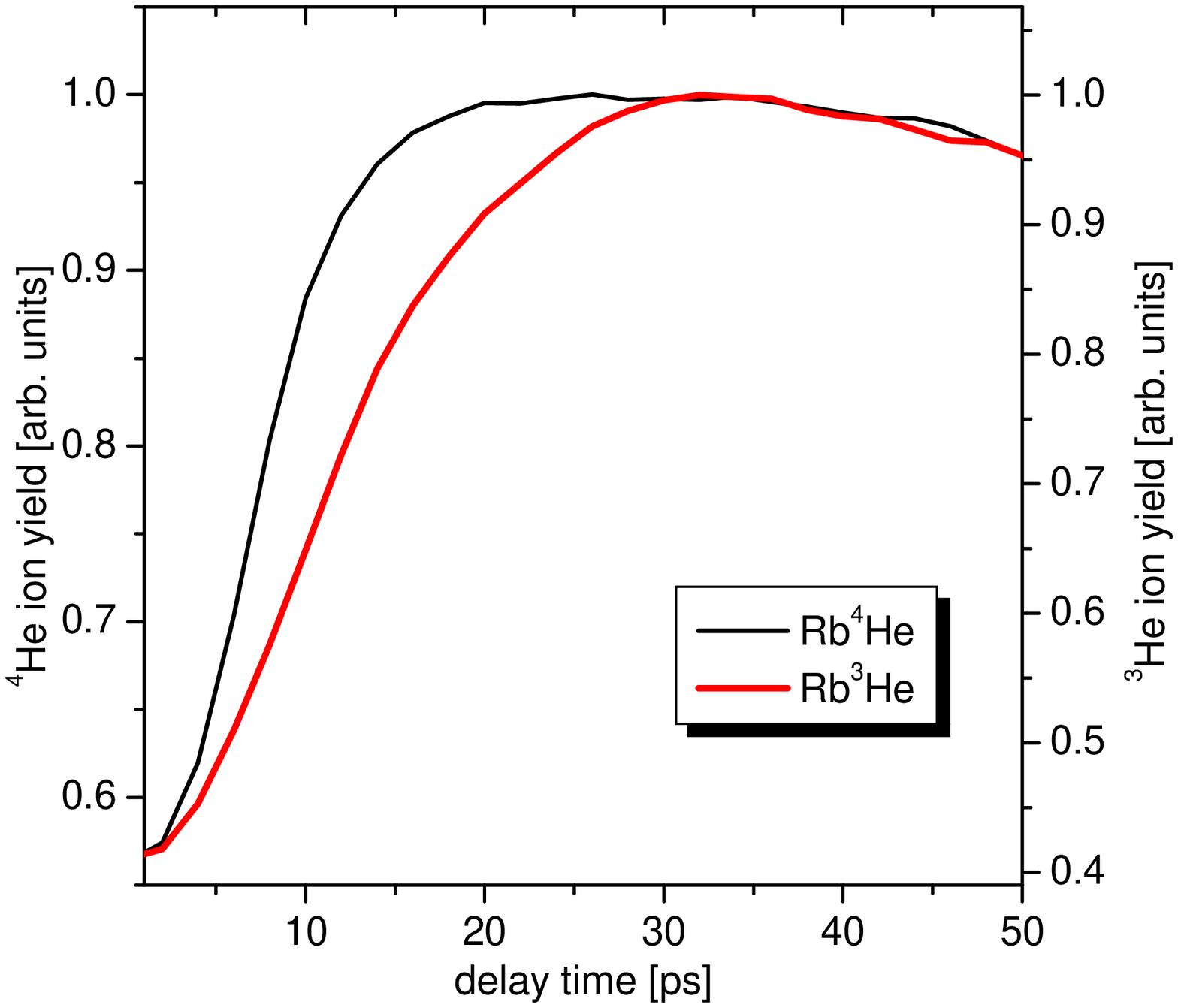}}
\caption{Femtosecond pump-probe spectrum exciting Rubidium atoms located at the surface
of helium nanodroplets. Plotted is the yield of RbHe molecules which form upon laser
excitation. Different formation times are observed for different helium isotopes.
Surprisingly the population of Rb$^4$He builds up in significantly shorter times
\cite{Sti:2004b}. } \label{fig:exciplexformation}
\end{figure}

The experiments of alkali atoms on the surface of helium droplets revealed a variety of
aspects on dynamical processes triggered by laser excitations. In several studies the
formation of alkali-helium exciplexes upon excitation of atomic transitions explained the
observed features in wavelength dispersed, laser-induced fluorescence spectra
\cite{Sti:1996a,Sti:1998b,Bruhl:2001b,Lehmann:2000c,Sti:2004b}. Fig.~\ref{fig:exciplexes}
illustrates the formation process. The alkali excited in the $p$-state will strongly bind
a He atom if it approaches in the nodal plane of the $p$ orbital. Hence an excited
alkali-helium molecule is formed upon $p\leftarrow s$ laser excitation, bound by a couple
of hundreds of cm$^{-1}$, depending on the corresponding alkali. Since the binding energy
of the exciplex to the droplet surface is weak, the binding energy release leads to
desorption. Eventually the complex dissociates when radiatively decaying into the
electronic ground state, typically after nanoseconds. To date, this process has been
experimentally confirmed for Na, K, Rb and Cs. Besides diatomic exciplexes one also
observes larger molecules (M$^*$He$_n$, M=metal, $n>1$) as known from studies in bulk
helium \cite{Sti:2001d,Nettels:2005}. From the beginning, the dynamics of this process
appeared to be an interesting example of photo-induced molecule formation. The first
experiment giving information on the dynamics of the formation process was carried out in
Princeton, applying time correlated photon counting \cite{Lehmann:1997,Lehmann:2000d}.
The evolution of the fluorescence of Na*He as well as K*He was monitored. A substantial
difference in formation rates was observed following excitation of the two different
spin-orbit component of the excited $^2$P alkali state. Modeling the data based on a
tunneling process into the molecular alkali-helium well, the contributions of the
extraction energy of helium atoms from the droplet and spin-orbit coupling effects were
discussed. Predicted exciplex formation times of the heavier alkalies were derived from
the model. Since the measured formation times of 50\,ps (K*He) approached the resolution
of the applied experimental technique, the use of femtosecond pump-probe techniques
became evident. Indeed, the formation times could be determined for K*He as well as Rb*He
employing mass-resolved multi photon ionization in a pump-probe arrangement
\cite{Sti:2000d,Sti:2004b}. These experiments included a direct comparison of $^4$He and
$^3$He droplets. Fig.~\ref{fig:exciplexformation} shows the real-time spectra monitoring
the yield of Rb*$^4$He and Rb*$^3$He. Formation times are determined to be 8.5\,ps  and
11.6\,ps, respectively. For both isotopes the exciplex formation occurs very much in the
same way, although we are dealing with very different fluids, on one hand  a
Bose-Einstein condensed $^4$He superfluid liquid, and, on the other hand the $^3$He Fermi
normal fluid. The longer formation time of the $^3$He exciplex is a very surprising
result and does not agree with the predictions of the  just mentioned tunneling model.
Different vibrational relaxation times are suggested to account for the difference, which
still has to be confirmed in a quantitative theoretical model.

In the same exciplex formation process quantum interference structures have been observed
\cite{Sti:1999b,Sti:unp_qi}. Recent experiments having attosecond time resolution in
recording interference fringes demonstrate that quantum interference structures are
suitable to get detailed information on the vibrational structure of the exciplex
molecules. Moreover, the interaction potential can even be monitored during the formation
of the metastable molecules. The interference oscillations survive the bond formation
process and provide a new tool to determine the energy of vibrational states with a
resolution on the order of $\approx 1$\,cm$^{-1}$ \cite{Sti:unp_qi}.

\subsection{Fragmentation of dopants}
Employing intense femtosecond laser pulses, femtosecond ionization of magnesium clusters
has been studied by high resolution mass spectrometry in the group of Meiwes-Broer
\cite{Doppner:2001}. Decomposition of the Mg-clusters dominates the process. The charging
of the fragments is significantly governed by the interaction with the helium droplet.
Pump-probe experiments on the MgHe$^+_N$ snowball formation clearly show dynamics up to
50\,ps.

Fragmentation dynamics upon ionization of neon clusters (Ne$_n$, $n<14$) embedded in
helium nanodroplets have been theoretically investigated by Halberstadt and coworkers
\cite{Bonhommeau:2004}. Here a Molecular Dynamics with Quantum Transitions (MDQT)
approach has been used; the helium environment is modeled in terms of friction forces.
Besides fitting the friction coefficient of helium droplets which surprisingly comes out
rather high compared to superfluid helium, fragmentation branching ratios are calculated
which are compared with experimental studies on electron impact ionization of neon-doped
droplets \cite{Ruchti:1998}. Here the helium environment significantly stabilizes larger
fragments. The calculations give a detailed view on the short time evolution of the
proportion of the species involved in the dissociation process.

\subsection{Photo-dissociation in helium droplets}
The first photo-dissociation experiments using helium droplet isolation were performed on
K triplet dimers and on Na and K triplet trimers, using both the methods of wavelength
resolved emission and time correlated photon counting. This work revealed that excitation
of K$_2$ to a largely repulsive $^3\Pi_g$ state leads to both the expected atomic K
emission and also to emission from molecules in the $^1\Pi_u$ state, demonstrating that
intersystem crossing in excited alkali states can compete even with direct
dissociation~\cite{Reho:2001}. In the case of the quartet trimers, the rate of
predissociation following electronic excitation, due to curve crossing by a repulsive
single surface, has been measured as a function of vibrational energy in the excited
state~\cite{Reho:2001a}.  This dependence suggests that the crossing occurs slightly
above the minimum in the excited state. The atomic and dimer emission to higher energy
than the excitation wavelength is allowed due to the formation of the stronger singlet
dimer bond upon dissociation of the weakly bound quartet trimer. The presence of dimer
emission from the $^1\Pi_u$ state but not the lower energy $^1\Sigma_u$ state points to
dissociation on a state that is antisymmetric with respect to the molecular plane of the
trimer.

\begin{figure}
\center\resizebox{0.8\columnwidth}{!}{\includegraphics{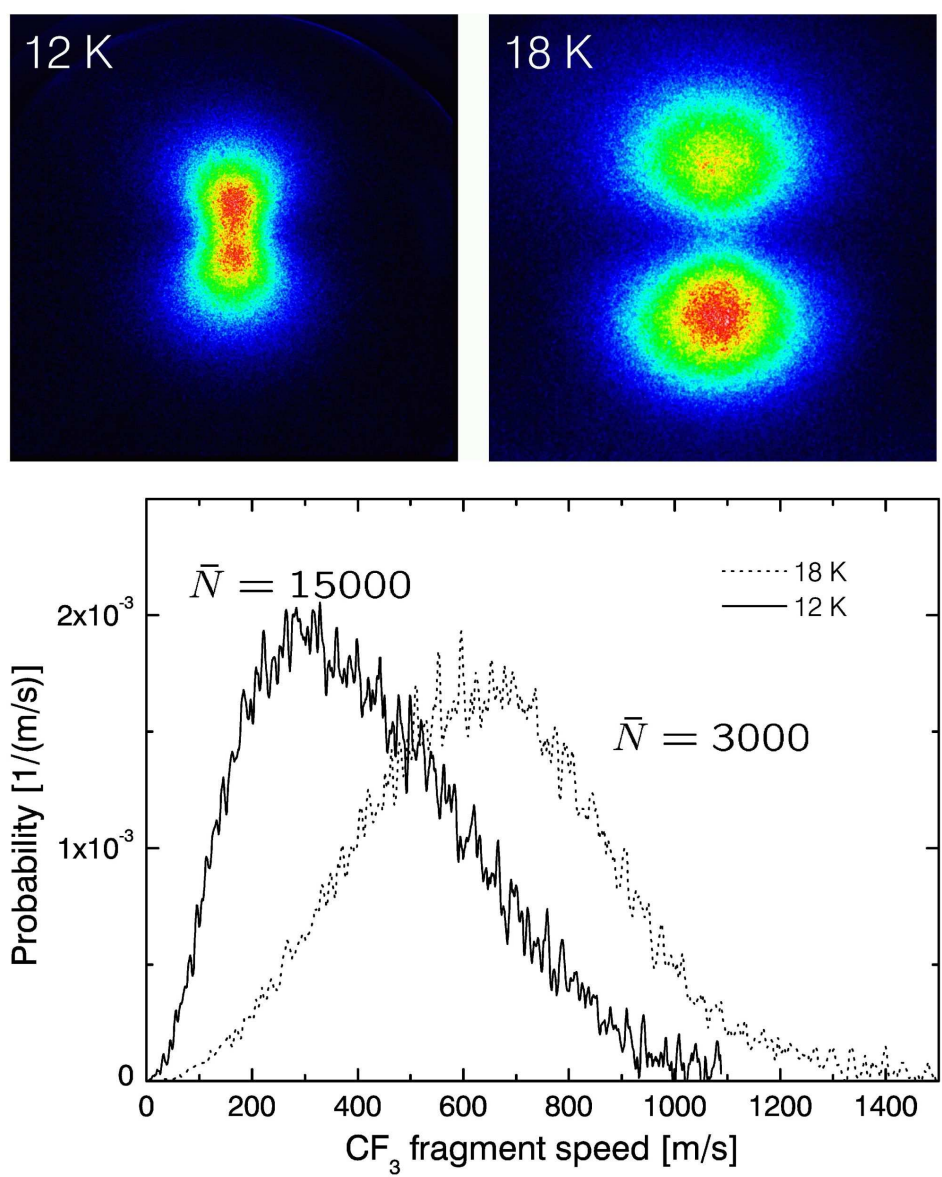}} \caption{Velocity
of CF$_3$ photo-fragments after photo-excitation of CH$_3$I for two helium droplet sizes
\cite{Braun:2004a}. The given temperatures indicate the nozzle temperatures used: 18\,K
corresponds to $\bar{N} \approx 3000$, 12\,K  to $\bar{N} \approx 15000$, respectively.
The velocity distributions (lower panel) were obtained by performing numerical Abel
inversions on the intensity distributions shown above.} \label{fig:Drabbelsphotolysis}
\end{figure}

A novel experimental strategy to directly probe the translational dynamics of neutral
species embedded in helium nanodroplets has been pursued by Drabbels and coworkers
\cite{Braun:2004,Braun:2004a}. They create fragments from a photo-dissociation process
with well-defined velocity distributions inside a helium nanodroplet. The comparison of
the fragments' initial and final (after having left the droplet) velocity distribution
provides detailed insight into the translational dynamics and the interaction with the
helium environment. The three-dimensional speed and angular distributions of various
departing reaction products are measured using a velocity map imaging setup.
Photo-fragments are nonresonantly ionized by femtosecond laser pulses, delayed 20\,ns
after triggering the photolysis. The 266\,nm A-band photo-dissociation of CH$_3$I and
CF$_3$I has been investigated inside $^4$He droplets having sizes from 2000 to 20000. For
all the examined sizes, some of the photo-fragments escape from the helium droplets,
though sometimes with a small helium cluster around them. Compared to classical solid
clusters where already a few solvation shells around the parent molecule lead to complete
caging \cite{Alexander:1988,Buck:2002,Vorsa:1996}, the helium droplets reveal
extraordinary dynamical properties. The measured velocities are found to be considerably
shifted to lower speeds with respect to the photo-dissociation of gas-phase
photo-dissociated molecules. Based on the observed speed distributions and anisotropy
parameters it is concluded that the CF$_3$ fragments escape via a direct mechanism, only
partially transferring their excess kinetic energy to the droplet. Iodine atoms, despite
their lower initial kinetic energy, escape with lower fractional loss of kinetic energy.
Accompanying Monte Carlo simulations suggest that mainly binary classical collision with
the helium atoms can account for the findings. Since recoil energies are quite high, no
quantum nature in terms of superfluidity and Landau's critical velocity had to be
introduced to model the findings. The low energy fragments show an isotropic angular
distribution. With increasing  recoil energy and product mass, the reaction product
angular distribution become similar to the free molecule. As expected, larger mean
droplet sizes result in higher kinetic energy losses and more isotropic angular
distributions of the departing products.  As an example, in
Fig.~\ref{fig:Drabbelsphotolysis} the angular as well as velocity distributions of the
CF$_3$ fragments dissociating CF$_3$I are shown. The molecules were dissociated using
pulses from the 266\,nm fourth harmonic of a Nd:YAG laser. The fragments were ionized in
a non-state-selective way employing femtosecond 800\,nm laser pulses delayed 18\,ns with
respect to the dissociation. Results obtained with different cluster sizes are compared.

\subsection{Photo-electrons of pure and doped helium droplets}
Many detection schemes include ionization processes of the embedded atoms and molecules.
The outcome and interpretation of results often is linked to the dynamics and the fate of
the produced charged particles. It is therefore instructive to study the produced
electrons by means of photo-electron spectroscopy. The first task in that direction
indicated that the kinetic energy distributions of electrons is dominated by slow
electrons having average energies less than 0.6\,meV, when photon energies of about
24\,eV are used for ionization of the droplets \cite{Peterka:2003}. A corresponding
strong interaction or ``thermalization'' of electrons produced  inside the droplet would
therefore limit the usefulness of photo-electron spectroscopic techniques for the study
of embedded species. Fortunately the results introduced in the following confirm that
this is not as a universal behavior following intra-droplet ionization.

The first photo-electron spectrum of an embedded neutral species was presented by
Tiggesb{\"a}umker and coworkers \cite{Radcliffe:2004}. Mass-selected Ag$_8$ clusters were
resonantly (R2PI) ionized by 3.96\,eV  photons. The ionization threshold was found to be
in good agreement with available theoretical calculations and previous gas-phase
experiments. In contrast to the experiments on the pure droplets, the ionization
threshold is not significantly altered by the helium environment. This is perhaps
surprising given that the conduction band for electrons in bulk helium is almost 1\,eV
above vacuum, which would appear to imply that the ionization threshold in helium should
be raised by nearly this amount.  In this regard the nanometer dimensions of the helium
droplets are crucial in the process ablating the electron and not forming a solvated
bubble state. Interestingly, the results show that prior to the absorption of the second
photon the system relaxes to a metastable state, most likely by vibrational relaxation.
Depending on the photon energy up to 80\,meV of energy are to be dissipated into the
helium droplet on a time scale much less than a nanosecond. A quite similar result was
found when studying the silver trimer \cite{Przystawik:2005}. Again, excess excitation
energy relaxes into a long living state, the lifetime of which was determined to be
$5.7\pm0.6$\,ns. The ionization potential was found to be in accordance with gas-phase
experiments; hence direct escape of the electrons from the droplet without significant
interaction appears to be the major channel.

Drabbels and coworkers recently studied photo-electron spectra of aniline attached to
helium nanodroplets \cite{Loginov:2005}. Although the spectra resemble closely that of
the gas-phase, a droplet size-dependent shift was observed, lowering the ionization
threshold upon solvation in the droplets. This shift, which is of the order of
800\,cm$^{-1}$, is assigned to polarization effects and can be readily estimated from the
dielectric constant and the cluster radius. In terms of resolution the photo-electron
peaks are asymmetrically broadened, showing a tail extending 100 -- 300\,cm$^{-1}$
towards lower kinetic energy. By analyzing the lineshape, droplet size dependent
contributions could be extracted. The results give a linear increase with the droplet
radius by $9.8\pm0.8$\,\,cm$^{-1}$\AA$^{-1}$. In conclusion, the line broadenings due to
the surrounding helium droplets are comparable to achievable width given by a common
apparatus function and are not a severe limitation in applying photo-electron
spectroscopy in helium nanodroplets.

\subsection{Theory of electrons attached to helium droplets}
The already mentioned experimental studies involve ionization processes. In particular in
the case of photo-electron spectroscopy the interaction of the emitted electrons with the
droplet importantly impact the results of the experiments. Jortner and Rosenblit
provided a thorough theoretical study of the energetics and dynamics of electron bubbles in
$^4$He and $^3$He droplets \cite{Jortner:2005}. The energetic stability is determined by
comparing the results to the energy of the quasifree electron state $V_0$ with the total
energy of the electron bubble in the ground state $E_t$. The latter includes
contributions of the electronic and reorganization energy. At a cluster size of 6500,
$V_0$ was obtained to be reduced only by less than 10\% compared to the bulk value of
$V_0=1.06$\,eV. Decreasing the droplet size rises the total energy $E_t$ of the
balloon-like structure continuously. The minimal droplet size for which an electron
bubble is energetically stable was determined to about $N = 5200$. For such droplets
having a radius of $39$\,{\AA} the radius of the electron bubble is $13.5$\,\AA. An
interesting result to point out here is the role of superfluidity. The energetic
structure proved to be insensitive to the properties of the superfluid, being nearly
identical for $^4$He below or above the lambda point as well as $^3$He droplets. Also the
calculated formation time $\tau_f$ of $9$\,ps of an equilibrium electron bubble
configuration is only weakly affected by superfluidity: dissipating effects in
normalfluid droplets increase $\tau_f$ by 15\%.

On the other hand, the dynamic stability, given by tunneling of the electron into the
vacuum is strongly affected by the superfluid phase. In normalfluid helium the motion in
the confining potential of the droplet is highly dissipative and therefore the electron
bubble rests in the center of the droplet~\cite{Lehmann:1999b}. In contrast to that, the
nondissipative motion in a superfluid droplet is predicted to decreases the tunneling
lifetime by 20 orders of magnitudes. Apparently $\tau_f$ strongly increases with droplet
size. Finally, defining a dynamic lifetime by $\tau_f$ being longer than the time scale
of an experiment (10$^{-6}$\,s), one obtains a minimal droplet size for dynamic stability
of $\approx 6200$. The theoretical picture and the numerical results nicely go along with
measurements on attached electrons to helium droplets and experimentally determined
lifetimes \cite{Farnik:1999,Farnik:2003}.

Rosenbilt and Jortner have also analyzed the binding of a free electron to the outer
surface of helium droplets.  Above a bulk helium surface, such an electron is weakly
bound by polarization forces.   However, due to curvature of the surface in droplets, the
binding energy is expected to be depressed, and a minimum droplet size for a bound
electron state is predicted to be 3 x 10$^5$ helium atoms~\cite{Rosenbilt:1994b}. Such a
threshold is not expected if the droplet contained a positive charge. The hypothetical
Rydberg type state (with principle quantum number $\ge$\,10) of an electron circulating
around a positively charged droplet, but prevented from neutralization by the helium
electron repulsion, has been named "Scolium".

\subsection{High resolution electronic spectroscopy}
High resolution electronic spectroscopy of doped helium nanodroplets can serve as a
powerful tool not only for obtaining structural information but also to study dynamics in
photo-chemical processes. In particular, charge transfer or proton transfer in complex
systems can be targeted. Often such processes take place in the femtosecond time regime
and real-time pump-probe techniques have been applied in ambient environments
\cite{Lippert:2003,Sobolewski:2002,Nibbering:2000,Kim:1995a}. Since lifetime effects in
this domain significantly contribute to line broadenings, frequency domain spectroscopy
can be applied as an alternative approach having advantages e.g.~in terms of sensitivity.
Limitations lie in the achievable spectral resolution and the assignment of spectral
features. In complex systems this can only be accomplished in a homogeneous cold ensemble
of molecules. Moreover, the efficient cooling mechanism present in helium nanodroplets
allows not only to state selectively laser-excited molecules but also to observe emission
spectra of well-defined states because vibrational excitations are cooled prior to the
radiation decay. The benefit of high spectral resolution in absorption spectra of complex
molecules has been demonstrated in a number of experiments
\cite{Lindinger:1999,Sti:2001e,Sti:2004a,Lehnig:2004c}. In some molecules, the first
vibrationally resolved electronic spectra have been recorded \cite{Sti:2004a,Sti:2005}.
This line of work has been extended to high-resolution fluorecence emission spectroscopy
by Slenczka and coworkers \cite{Lehnig:2003,Lehnig:2004a,Lehnig:2005}. Molecules like
tetracene, pentacene, perylene and phthalocyanines have been studied. The results clearly
demonstrate discrete and long lived states of the solvation structure of the surrounding
helium matrix. Explicitly, the emission spectra show contributions from different
progressions. Since vibrational modes of localized helium atoms are not expected to exist
in superfluid helium, the experiments confirm the existence of a solid like (snowball)
solvation shell \cite{Lehnig:2004a,Lehnig:2005}. Depending on the molecule, different
helium layer configurations have been assigned and relaxation probabilities were derived.
QMC calculations have found two different helium solvation structures for 150 He atoms
around phthalocyanine \cite{Whitley:2005}. For pentacene e.g.,the authors found only a
single configuration of the helium layer, independently on the electronic state. More
generally, multiple sharp lines are often observed for each vibronic transition of
polyatomic molecules \cite{Hartmann:1998, Lindinger:2001a, Sti:2000d}, along with a broad
absorption feature to the blue of the lowest energy sharp transition.   The broad peak
has been assigned as a ``phonon wing'' involving excitation of bulk like phonons and
rotons in the droplet \cite{Hartmann:1996a}. The additional sharp lines have been
assigned in some cases as excitations of helium vibrations localized in the solvation
shell \cite{Hartmann:2002, Lehnig:2004c} or as zero phonon lines (pure molecular
excitations) of alternative helium structures. The intensity of the zero phonon line is
proportional to the squared overlap of the ground state helium wavefunction solvated
around the ground and the excited state molecule respectively, and thus is expected to
decrease extremely rapidly with increasing solvent reorganization. Hole burning
experiments  as well as dispersed emission as described above can be used to
assign peaks to different solvation structures (as was first done in the spectrum of
tetracene \cite{Hartmann:2001, Lindinger:2001b}), but this requires the structures to
have a lifetime longer than the excitation laser pulses.  It is worth noting that the
localized vibrations and different isomers can be viewed as points on a continuum.   As
helium density in the first solvent shell increases, one or more helium mode become
``soft'' and move down in excitation energy below the roton energy (where it belongs
based upon its wavelength). As the density further increases, the ``soft'' mode can
become an unstable mode with multiple minima.  The later can be viewed as a result of
freezing of the highly compacted first solvation shell due to strong He--He repulsion.

Generally, the resolution of electronic spectra has not been sufficiently high to allow
the resolution of rotational structure.   An exception is glyoxal \cite{Portner:2002} for
which a surprisingly large change in rotational constants upon electronic excitation was
observed an attributed to be primarily due to changes in helium solvation structure. The
zero phonon spectrum of phthalocyanine was studied with 1 MHz resolution and found not to
have any evidence of rotational structure \cite{Slenczka:2001}.  This work determined
that the experimental width of the zero phonon line was dominated by the inhomogeneous
size distribution of the droplets, even for the largest droplets that could be formed.
The asymptotic shift with droplet size is expected to be proportional to the droplet
helium number and the change in helium-molecule C$_6$ coefficient. The latter is
typically much larger for electronic than vibrational excitation, thus making droplet
size inhomogeneous effects more important in electronic than vibrational spectroscopy
\cite{Dick:2001}.

HENDI has also been used to study the electronic spectroscopy of complexes formed inside
the helium droplets.  Additional rare gas atoms such as Ar are often used
\cite{Hartmann:1998}, since it is common to have jet spectra of such complexes to compare
to. Recent examples include the study of tetracene with H$_2$, D$_2$ (ortho and para
forms), and HD \cite{Lindinger:2004}.  They discovered multiple isomers of each complex,
some of which were not observed in free jet spectroscopy of the same complexes. It was
recently observed that the beam depletion spectrum of perylene when complexed with O$_2$
was much stronger than expected based upon the Poisson distribution and this was
attributed to increased quenching and thus greater heat release per excitation
\cite{Carcabal:2004}. This suggests that comparison of LIF and depletion spectra could be
used to determine quantum yields for emission in helium.  A recent study reported the UV
spectrum of benzene and the benzene dimer in helium \cite{Schmied:2004}. The monomer was
found to have a larger blue shift (30\,cm$^{-1}$). In a more recent measurements on five
substituted benzene molecules, it was found that their blue shifts (and that of benzene)
was highly correlated with calculated changes in the electron density on the aromatic
ring \cite{Boatwright:2005}. Curtis \textit{et al.} \cite{Curtis:2005} found a $\approx
1$\,cm$^{-1}$ blue shift for the benzene transitions when they are observed using two
photon ionization detection compared to detection by LIF or beam depletion
\cite{Schmied:2004}.  It is likely that this reflects the REMPI detection being sensitive
to the detection of smaller droplets for which the electron has a higher probability of
escaping. The benzene dimer was found to have the same spectroscopic structure as
previously found in the gas phase, but compressed \cite{Schmied:2004}.  Remarkably,
helium solvation was found to substantially inhibit excimer formation upon S$_1$
excitation, a process that occurs in a few psec for the isolated dimer in the gas phase.
Krasnokutski \textit{et al.} \cite{Krasnokutski:2005} have reported the spectrum of
anthracene in helium, thus completing the observation of the series of one to five fused
benzene rings in a line. Huang \textit{et al.} \cite{Huang:2004} have done QMC
calculations of the interaction of helium (up to 24 atoms) with these ``nanosurfaces''
and find strong localization of helium in the first solvent layer above and below the
rings. Pendular state spectroscopy has been observed for 9-cyanoanthracene in helium
using fields up to 200 kV/cm, achieving what the authors report to be the highest degree
of alignment yet obtained \cite{Kanya:2004}. It is likely that such spectra could provide
measurements of the moments of inertia of such large molecules in helium since the
pendular frequencies will depend upon these moments.

One of our groups (FS) introduced the possibility of studying charge transfer and life
time effects in doped helium nanodroplets when probing
e.g.~3,4,9,10-perylenetetracarboxylic-dianhydrid (PTCDA) complexes
\cite{Sti:2003,Sti:2005}. Only the high selectivity of molecular transitions attained in
helium droplets allows for a detailed analysis on excitonic transitions and a
determination of their life times. Evaluating the spectral widths, a dephasing time of
ca.~10\,fs was deduced. Analysis of line shapes in 3-hydroxyflavon have been performed by
Slenczka and coworkers \cite{Lehnig:2004}. They determined the proton back transfer of
tautomers of 3-hydroxyflavon into its normal form in 250\,fs. Furthermore, the influence
of a polar solvent environment by adding water molecules was probed. Note that these
experiments have full control on the size of the additional water complexes.

\subsection{Optically selected mass spectroscopy in helium nanodroplets}
There is a long history to the study of electron impact ionization of helium
nanodroplets, of which we cite only some of the most recent \cite{Callicoatt:1996,
Callicoatt:1998a}. The Miller group introduced a new approach that dramatically cleans up
many of the ambiguities of previous work, by looking only at the modulation in the mass
spectrum induced when a IR transition of a particular species is
pumped~\cite{Lewis:2005}. This allowed, for the first time, reliable absolute branching
ratios between different mass (ion products) to be determined, from droplets with a known
composition. The work refined the ion hopping probability, and demonstrated that there is
a long range steering, such that the He cation ion preferentially approaches the negative
parts of the molecule, leading to regio-selective ionization. Fragmentation after
ionization was studied for HCN \cite{Lewis:2005a} and the large organic molecule,
triphenyl-methanol \cite{Lewis:2004}.

\subsection{IR-IR double resonance}
The high power available from OPO's has been exploited by the Miller group to use IR-IR
double resonance to study dynamical behavior in helium nanodroplets. The C-H stretching
spectrum of cyanoacetylene (HCCCN) was studied as a test case~\cite{Merritt:2004}.   A
high power 3 micron OPO was used as a pump and a lower power F-center laser used as a
tunable probe of the spectrum $\approx 175\,\mu$s after the the pump, sufficient time to
allow evaporative cooling to effectively finish.   As expected, a saturation hole in the
probe absorption was induced by the pump.   However, this was narrower than the one
photon transition and an increase in absorption (a ``hill') was observed slightly higher
in wavenumber. The interpretation is that the cluster size distribution makes a
significant contribution to the width of the single resonance spectrum, and that there is
a blue shift of the absorption spectrum due to the evaporation of $\approx 660$ helium
atoms induced by the pump laser absorption. It should also be recognized that the total
angular momentum will be conserved in the droplets and that the initial absorption will
lead to a reduction in the angular momentum trapped in the droplets following initial
pickup, so that change in the droplet spectrum may not reflect only the change in mean
size.

In a subsequent study, Miller and coworkers used double resonance to study the IR induced
isomerization between the linear HCN--HF and the bent HF--HCN
complexes~\cite{Douberly:2005a}.   Both complexes are formed upon joint pickup of HCN and
HF~\cite{Douberly:2005b}, and in each isomer both the C-H and F-H stretching modes were
observed.   Excitation of the lower energy HCN--HF complex resulted in branching
probabilities of 58\% to produce HCN--HF absorption in a smaller droplet and 29\% to
produce absorption of the higher energy HF--HCN isomer.   The results are interpreted to
result from photo-dissociation of the complex after IR absorption (which is commonly
observed for isolated complexes) followed by recombination, much as in the initial
formation process of the complex following pickup.  The $\approx 13$\% loss of intensity
was attributed to droplets lost from the detector due to the increased transverse
momentum of the droplets produced by the evaporative cooling.   In particular, no
evidence was found for an IR induced absorption of the isolated HCN or HF molecules in
the droplets, which implies that ejection of one of the monomers due to the translational
energy imparted upon photo-dissociation is at most a minor channel.   Pumping the
vibrations of the higher energy HF--HCN isomer, in contrast, lead to quantitative
transfer of the population to the lower energy HCN-HF isomer.  Two mechanisms where
proposed to explain this result.   One is that for this complex, vibrational energy
redistribution (IVR) is sufficiently fast that photo-dissociation does not take place.
Instead, one creates a vibrationally hot molecule with energy well above the
isomerization barrier, and that this cools sufficiently slowly that the system anneals to
the lowest minimum structure.  The alternative is that photo-dissociation does take place
but produces a vibrationally excited product that does not relax until after
recombination.   The subsequent vibrational relaxation of the complex then leads to
annealing.  Time resolved pump-probe spectroscopy, with sufficient time resolution to
`catch' the absorption spectrum of the fragments before geminal recombination is clearly
needed to resolve this ambiguity in the mechanism.

\subsection{Complex formation in helium droplets}
Helium nanodroplet spectroscopy has been used to study complexes since the first IR
spectral study by Goyal \textit{et al.} \cite{Goyal:1992b}.   Here, we will restrict
discussion to work that have appeared since the entire field was reviewed in the special
issue  on helium nanodroplets that appeared in the Journal of Chemical Physics in late
2001.

The study of weakly bound complexes formed in supersonic expansions is a well established
method that has provided great insight into noncovalent interactions between molecules.
Likewise, study of the IR spectra of complexes trapped in cryogenic matrices,
particularly Ar and Ne, is also a well established technique. Study of complexes in
helium offers several features that compliment these other methods. For one, the long
range ``steering'' of molecules when formed in helium can lead to the selective
production of structures that are not formed significantly in jets or classical matrices.
The cooling provided by the helium allows the quenching of complexes in higher energy
structures that are separated by even modest barriers from the global minimum structure.
The most dramatic example of this is the formation of long, polar complexes of HCN and
HCCCN \cite{Nauta:1999, Nauta:1999a}.  A more recent example is the exclusive formation
of a ``open'' polar form of formic acid dimer \cite{Madeja:2004}. Such barriers are
particularly important when preformed hydrogen bonded rings must be broken up to reach
the minimum energy structure for the next larger complex. In a study of the growth of HF
polymers in helium, it was found that a cyclic tetramer is formed (which requires
insertion into the cyclic trimer), but that the fifth HF is not able to enter the ring
and thus leads to a polar pentamer \cite{Douberly:2003a}.   The formation of cyclic water
complexes has also been studied \cite{Nauta:2000a, Burnham:2002}. For the complex of HF
and HCN, both hydrogen bonded complexes are formed in helium \cite{Douberly:2005b} while
only the lower energy HCN--HF complex is observed in jet spectroscopy. In contrast to
that, for complexes between HCCH and HF only the ``T'' shaped isomer is observed in
helium despite a predicted minimum in the HCCH-FH structure \cite{Douberly:2003a}. It is
presumed that the later is shallow enough that the system can tunnel out of this minimum
even if initially populated.   It has been demonstrated that helium can trap and allow
the study of ``entrance channel complexes'' (Cl, Br, I)--HF \cite{Merritt:2005}.

Helium nanodroplets are particularly suited for the study of molecules bound to small
metal clusters grown inside the droplet.   The Miller group has published several papers
reporting the IR spectra of HCN \cite{Nauta:2001g, Stiles:2003a, Stiles:2004}, HCCCN
\cite{Dong:2004b}, and HCCH \cite{Moore:2004b} complexed with Mg atoms and small
clusters.   Strongly nonadditive shifts in the IR fundamental transition of the molecules
are suggestive to changes in the bonding of the Mg clusters, particularly upon going from
Mg$_3$ to Mg$_4$ \cite{Stiles:2004}.

The Toennies-Vilesov group published an important series of papers on the IR spectra of
OCS complexed with different numbers of hydrogen and its isotopomers
\cite{Grebenev:2000d, Grebenev:2001a, Grebenev:2002, Grebenev:2003}.  A controversial
finding of this work is certain para-H$_2$ clusters become superfluid between the
temperatures of 0.37\,K (found in $^4$He droplets) and 0.15\,K (found in mixed
$^3$He/$^4$He droplets).   The principle observation supporting this interpretation is
the loss of Q branches in the spectrum of these complexes when upon cooling. Paesani
\textit{et al.} \cite{Paesani:2003b, Paesani:2005b} have performed theoretical
calculations that provide support for this interpretation. One of us (KL) and others have
argued that this observation is consistent with a set of pH$_2$ rings that can internally
rotate (cyclic exchange) and the expected spin statistical weights of the internal
rotation levels \cite{Callegari:2001}.  A later paper from Toennies and Vilesov
\cite{Grebenev:2002} reached substantially the same conclusion.  It is a semantic
question whether such 1-D rotations should be considered as ``superfluid''.

The Miller group has studied HCN \cite{Moore:2001a,Moore:2001b,Moore:2003a,Moore:2003b}
and HF \cite{Moore:2004a} complexed with ortho and para H$_2$ and D$_2$, and with HD,
including mixed clusters. Perhaps the most important result of this work is the study of
HCN-(HD)$_n$ complexes, where  Q branches were found to disappear for $n$ = 12 and 13 and
reappear for $n$ = 14 \cite{Moore:2003b}. This was interpreted as due to the formation of
cages which nearly isotropic potentials for HCN rotation for the n=12 and 13 cases. Since
HD is a composite Fermion, it is clear that the disappearance of Q branches in the IR
spectrum cannot be viewed as conclusive experimental evidence for formation of superfluid
solvation shells, no matter how liberally one wants to interpret that phase.

\subsection{IR spectra of isolated molecules}
The bulk of more recent IR studies of isolated molecules have focused on probing the
interactions between the molecule and the helium solvent and the dynamics that are
produced.   Several studies have explored enhanced line broadening in IR spectra that
reflect rapid, solvent induced relaxation, often facilitated by intramolecular anharmonic
resonances.    Madeja \textit{et al.} studied the spectrum of h(2)- and d(1)- Formic Acid
in the spectral region of the O-H and C-H stretching vibrations, as well as several
combination bands that gain intensity through Fermi and Coriolis resonances
\cite{Madeja:2002}.  Lindsay and Miller studied the C-H stretching fundamental spectrum
of ethylene \cite{Lindsay:2005}, while Scheele \textit{et al.} studied the first C-H
stretching overtone region of the same molecule \cite{Scheele:2005}.  This work again
demonstrated extreme variation in broadening between different vibrational bands.
Slipchenko and Vilesov have re-examined the 3 micron spectrum of NH$_3$
\cite{Slipchenko:2005}, greatly extending a previous, lower resolution study
\cite{Behrens:1998}.  This work demonstrated that the rotational constants (B and C) are
reduced by only 5\% compared to the gas-phase values and that the inversion splitting is
within 6\% of the gas phase value.  The later is perhaps surprising in that one would
expect a significant change in helium solvation upon inversion, therebye increasing the
tunneling mass.

The Havenith group has studied the fundamental IR spectra of both NO
\cite{vonHaeften:2005b} and CO \cite{vonHaeften:2005c} in helium. In the case of NO, they
found that the rotational spacing between the lowest two levels was 76\% of the
corresponding gas phase value, demonstrating that this molecule is in the ``intermediate
following'' limit, as had previously been experimentally demonstrated for HCN and DCN
\cite{Conjusteau:2000}. For the Q(0.5) $^2\Pi_{1/2}$ transition (between the lowest
rotational levels of each vibrational state), the transitions are quite narrow and laser
limited ($\approx 150$\,MHz FWHM), and they were able to resolve both the $\Lambda$
doubling and hyperfine splittings.   The hyperfine interaction constant is unchanged in
helium, demonstrating the small effect of the helium on the electronic density of the NO
molecule.   The $\Lambda$ doubling constant is increased by 55\% compared to the gas
phase.   This is the opposite effect from that expected based upon the change in B value
and also the expected increase in the Rydberg type excited states.   These authors point
out that ~90\% of the $\Lambda$ doubling arises from mixing in of excited $^2\Sigma$
valence states, which are expected to be only weakly shifted in helium.   The authors
suggest that confinement of the unpaired electron by the helium may increase the matrix
element $< \Sigma | L_y | \Pi>$ (which appears in the expression for the $\Lambda$
doubling splitting). They offer as an additional possible source for the increased
$\Lambda$ doubling helium density fluctuations in the helium droplet that slightly
perturb the rotational symmetry around the NO axis.  For a rigid NO in a $^2\Pi$ state,
the helium density should undergo Jahn-Teller distortion leading to a difference in
helium density in the nodal plane of the $\Pi$ orbital compared to that perpendicular to
it, i.e. leading to the anisotropy of helium density that the authors predict. For the
Q(1.5) transition, the fine and $\Lambda$ doubling splittings are unresolved, and the
transition is Lorentzian with a FWHM of 1.05\,GHz, which suggests that the upper
rotational state relaxes with a population lifetime of 152\,ps. This is a relatively fast
decay (for helium).  Rotational relaxation $J' = 1.5 \rightarrow 0.5$ will liberate
4.965\,cm$^{-1}$ of energy, well below the energy required to produce a roton ($\approx
6$\,cm$^{-1}$ in bulk helium).  The authors attribute this broadening as due to coupling
of the NO rotation to droplet phonons.  As discussed above and also by the authors in
their CO paper (see below), one must assume that the phonons are themselves coupled to
some much higher density of states in order to rationalize the observed smooth Lorentzian
lineshape.

In the IR spectrum of CO, only the R(0) line could be observed due to the low temperature
of the droplets relative to the rotational spacing.  However, by observing the same
transition in four isotopic species ($^{12}$C$^{16}$O, $^{13}$C$^{16}$O $^{12}$C$^{18}$O,
and $^{13}$C$^{18}$O) and exploiting the different reduced mass dependence for the
respective constants, the shifts for both the vibrational frequency and rotational
constant (more properly the J = 0 to 1 spacing) were deduced. The rotational constant of
$^{12}$C$^{16}$O is 62.9\% of the gas phase value, confirming that CO is also an
intermediate following molecule. The authors point out that this reduction is larger (in
percent) than that of HCN (79\%), despite the fact that the gas phase rotational constant
of CO (1.922\,cm$^{-1}$) is larger than that of HCN (1.478\,cm$^{-1}$).  This is
attributed as due to a significantly higher anisotropy of the interaction potential of CO
with He compared to that of HCN with He. Quantum Monte Carlo calculations on clusters
CO-He$_N$ ($N$=1-30) by Cazzato \textit{et al.} \cite{Cazzato:2004} rotational excitation
energies combined with vibrational shift calculations by Paesani and Gianturco
\cite{Paesani:2002} are in excellent agreement with IR experiments on such cluster sizes
\cite{Tang:2003a}.  These calculations found that the rotational excitation energy was
almost constant after completion of the first helium solvation shell ($N$=14), and by
extrapolation predicted a nanodroplet B value 78\% of the gas phase value, considerably
larger than what was found \cite{vonHaeften:2005c}. von Haeften \textit{et al.} interpret
this extra reduction as due to interaction of the CO rotation with long wavelength
phonons in the droplet which are not present in the small clusters. This is supported by
calculations using the CBF/DMC method \cite{Zillich:2004a}.
These calculations predict a nanodroplet CO rotational excitation 69\% of the gas phase
value (compared with 63\% inferred from the experiment).

For all four isotopic species of CO, the R(0) transitions exhibit Lorentzian lineshapes
with FWHM of 1.02\,GHz, almost identical to that observed for the NO Q(1.5) transition.
In the CO case, the energy released upon rotational relaxation is only 2.2\,cm$^{-1}$,
well below the energy of both roton formation and quantum evaporation (4.4\,cm$^{-1}$ for
droplets of 2\,600 helium atoms).  The CBF/DMC calculations (which treat the phonons as
continuous instead of discrete as they are in nanodroplets) predict a homogeneous width
of 0.54 GHz, approximately one half the observed value.   The authors discuss the fact
that the phonons alone cannot provide the needed quasi-continuum to produce a
homogeneously broadened line of 1 GHz width. They report that if one includes all the
states generated by coupling of droplet phonons and center of mass motion states
\cite{Lehmann:1999} for the CO, this produces a density of about 10 states per linewidth
of the CO transition.  They assign the line broadening as arising from inelastic
relaxation of the excited J = 1 level of CO into J = 0 levels of CO with production of
excitation in mixed phonon/center of mass motion states.  No mechanism was proposed for
the coupling that would produce such simultaneous  excitation of phonons and
translational motion. We would like to note that such a coupling is higher order and thus
likely weaker than the direct coupling of the rotor to the phonons.  The center of mass
motion potential is nearly harmonic and thus strong propensity rules for its change in
quantum numbers are expected. We would also like to point out that elastic orientational
relaxation (changes in the M quantum number) will also lead to line broadening in the IR
spectrum and this is most effective for low $J$ states \cite{Rabitz:1970}. The ripplon
states excited in the droplets (and the translational motion states) have a high
degeneracy in the harmonic limit and these could couple to the angular momentum of the
molecule (as was explicitly calculated for HCN in \cite{Lehmann:1999}). Time resolved
pump/probe experiments with polarized light could provide unambiguous determination of
the relative contributions of elastic, inelastic, and pure dephasing contributions to the
homogeneous broadening in the spectrum of CO and other molecules.

Nauta and Miller reported the spectra of diacetylene and cyclopropane \cite{Nauta:2004}.
In both cases, the rotational constants are the factor of $\approx 3$ smaller than the
corresponding gas phase values expected for heavy rotors.   The asymmetric C-H stretch of
diacetylene shows a small vibrational origin red shift (-0.304\,cm$^{-1}$) in contrast to
the small blue shift (+0.13\,cm$^{-1}$) observed for the corresponding band of acetylene.
Rotational independent linebroading with FWHM of 1.6\,GHz is observed \footnote{Note that
there is a typographic error in the unit of the linewidth reported in Table 1 of this
paper, as inspection of the simulated spectrum presented in figure 1 makes obvious.  The
unit should read cm$^{-1}$, not MHz.} which is attributed to vibrational relaxation. For
the observed $\nu_8$ C-H stretching mode of cyclopropane, the fundamental wavenumber is
blue shifted by +4.069\,cm$^{-1}$, rather a large vibrational shift. K{\"u}pper
\textit{et al.} obtained the 3\,$\mu$m $\nu_1$ spectrum of propargyl radical, formed from
a pyrolysis source \cite{Kupper:2002}.  The authors provided the first experimental
determination of the dipole moment (-0.15\,Debye) of this important free radical species.
Stiles \textit{et al.} \cite{Stiles:2003b} carefully studied the Stark effect  to
determine the effective dipole moments in helium of HCN (2.949(6) compared to
3.01746\,Debye in gas phase) HCCCN (3.58(8) compared with 3.73172\,Debye for gas phase).
This work demonstrates for the first time the perturbation that helium spectroscopy
introduces into the determination of dipole moments.  The authors found that these  small
shifts in the effective dipole moments could be accounted for by helium polarization,
using an elliptical cavity model with realistic parameters.

The Vilesov group has used an OPO and H$_2$ crystal Raman shifting of the OPO to
reexamine the IR spectra of several simple molecules, including H$_2$O, NH$_3$, and
CO$_2$. In the case of H$_2$O \cite{Kuyanov:up}, they have found that all the monomer
transitions can be assigned (three from the $\nu_3$ band and two from the $\nu_1$ band)
start from the lowest ortho and para rotational levels.  Transitions previously assigned
to higher lying rotational levels of H$_2$O (which would imply incomplete rotational
relaxation) were established to be water dimer transitions.  From the limited number of
transitions, only one combination difference could formed (between the 2$_{02}$ and
0$_{00}$ levels in the $\nu_3$ excited state) and this differs by only 0.5\% from the
corresponding gas phase value, indicative of the tiny effect helium has on the rotational
constants of H$_2$O. The three transitions to levels that are symmetry allowed to undergo
rotational relaxation have FWHM's between 2.3-3.0\,cm$^{-1}$, which implies lifetimes of
2\,ps. The two transitions that go to the lowest ortho or para rotational level in the
excited vibrational state have nearly laser limited linewidths of 0.34\,cm$^{-1}$.
Saturation measurements indicate that the excited state vibrational relaxation lifetime
is at least 7\,ns, the length of the excitation pulse.

Slipchenko and Vilesov \cite{Slipchenko:2005} observed the 3\,$\mu$m spectrum of NH$_3$
and observed a total of ten transitions in the $\nu_1$, $\nu_3$, $2\nu_4$(l = 0), and
$2\nu_4$(l = 2) vibrational bands.  They also reassigned an earlier spectrum observed in
the $\nu_2$ umbrella fundamental. Like for H$_2$O, transitions only come from the lowest
rotational level for the ortho and para species, though the upper tunneling level of the
J = K = 1 level is weakly populated. All transitions are close to the gas phase values,
and fits (which require some constraints given the few transitions observable) predict B
and C rotational constants $\approx$95\% of the gas phase values, as expected for such a
light rotor (B = 9.96\,cm$^{-1}$). The sum of the tunneling splitting of the ground and
excited $\nu_1$ state was found to be 1.67(5) compared to a gas phase value of
1.78\,cm$^{-1}$, indicating a small perturbation on the inversion frequency, especially
compared to the previous analysis of the $\nu_2$ spectrum.  Again, as in the case of
H$_2$O, transitions to levels that have symmetry allowed rotational relaxation channels
are broadened, indicating lifetimes of 1-7\,ps, while transitions to levels with only
vibrationally inelastic decay channels have nearly laser limited widths of
0.33\,cm$^{-1}$\, FWHM. This includes the sQ(1,1) level which could relax via relaxation
by transition to the lower tunneling inversion level (a $\rightarrow$ s) in the excited
state.

Hoshina \textit{et al.} \cite{Hoshina:2005} recently reported observation of a ``phonon
wing'' in the R(0) absorption of  CO$_2$ in the fundamental band \cite{Hoshina:2005}.
All previous IR spectra were assigned to pure excitation of the solute molecule, but
observations of phonon wings are quite common in electronic spectra in helium.  The
strength of the phonon wing reflects changes in the helium solvation structure upon
excitation, which are expected to be much larger upon electronic than vibrational
excitation.   The large cross section of the CO$_2$ fundamental band combined with the
high peak power of the pulsed laser source used allowed the broad, phonon wing to be
brought up in intensity by saturation of the pure molecular excitation (zero phonon
line).  These authors further reported that the fractional intensity in the phonon side
band can be predicted using a ``Toy model'' of a rigid, planar ring of helium atoms
coupled to the rotation of the molecule \cite{Lehmann:2001}.  This suggests that in this
case that the phonon wing does not arise from the changes in helium solvation structure
upon vibrational excitation, but rather due the rotational excitation of the molecule and
its coupling to helium rotation of in the first solvent shell.   A prediction of this
model for the phonon sideband is that the relative strength should be very weakly
dependent upon the degree of vibrational excitation and also that the phonon sideband
should be predictably stronger for the R(1) transition, which could be observed in the
C$^{18}$O$_2$ isotopic species.

\subsection{Small helium clusters}

While outside the formal scope of this review, we would like to close by briefly
reporting on important recent work on helium clusters smaller than those we have
considered until now.

Toennies and coworkers have continued to study small $^4$He and mixed $^3$He/$^4$He
clusters using transmission diffraction gratings to spatially resolve clusters by mass as
they introduced in \cite{Schollkopf:1996}.  Recently, they reported enhanced production
of $^4$He clusters with certain ``magic number'' sizes ($N$ = 10-11, 14, 22, 26-27, and 44
atoms) \cite{Bruhl:2004}.  This was quite surprising as the best DMC calculations
\cite{Sola:2004} indicate that no enhanced stable clusters should exist based upon their
ground state energy.   However, the magic number clusters are predicted to have reduced
Free energy due to the creation of additional stable ripplon modes as the clusters grow
\cite{Bruhl:2004}.

Particularly germane to the subject of this review is a series of papers reporting the IR
(McKellar Group) and microwave spectra (J{\"a}ger group) of a simple molecules with growing
numbers of helium atoms.   These include OCS \cite{Tang:2002d, Tang:2003b, Xu:2003}, NNO
\cite{Xu:2003b}, CO$_2$ \cite{Tang:2004a, Tang:2004c}, and CO \cite{Tang:2003a,
McKellar:2004}.  They have been able to follow how the spectra evolve as helium atoms
increasingly solvate the molecules, following the spectra up to $N$ = 19 for the cases of
CO and NNO.  For the cases besides CO, the first helium atoms strongly bind to the side
of the linear molecules and a fairly rigid ring builds up.   The moments of inertia are
close to the classical values and raise with increasing number of helium atoms.
However, once this ring is saturated, the additional helium atoms become more loosely
bound, primarily on the ends of the molecules.  This opens up the possibility of quantum
exchange and leads to a decrease in the moment of inertia as more helium atoms are added.
In detail, the rotational constants have a complex and highly structured evolution with
number of helium atoms which provides a window into the changing rotational dynamics.  It
is clear that the convergence to the nanodroplet limit is quite slow with a lot of road
to be yet explored.   Particularly insightful have been a series of DMC calculations that
have extracted the excitation energies by fits to the imaginary time correlation function
for the rotor orientation \cite{Moroni:2003, Moroni:2004, Cazzato:2004, Paolini:2005,
Moroni:2005, Paesani:2004b, Paesani:2005a, Paesani:2005b}. These have been in
quantitative agreement with the experiments, nearly reproducing every detail of the
changes of rotational constant with $N$.

\section{Conclusions}
As we hope this review makes clear, the field of helium nanodroplet spectroscopy is
advancing rapidly, and much has been learned in the past years. The field is mature in
the sense that many of the basic spectroscopic properties of molecules solvated in helium
can now be predicted. The location and binding of a given solute at the surface or in the
bulk of the droplets is well conceived as well as the temperature determination and
cooling mechanisms. We have a good understanding of the size of helium perturbations to
the spectroscopic constants of the molecule, though we still lack generally predictive
theories. We have learned how to exploit the unique properties of helium droplets to make
novel chemical species and probe them in new ways. However, our understanding of the
dynamical coupling of the solute to the helium remains limited. Ultrafast pump-probe
experiments have provided the first direct view into the reorganization of the helium
solvent following photo-excitation. Moreover, several recent pump-probe experiments done
with nsec and even cw lasers have also provided windows into the dynamics of molecules in
helium on times scales longer than accessible with ultrashort pulses. The feasibility to
study complex dynamical processes is demonstrated in the mentioned experiments, but much
more can and needs to be done. In particular, we look forward towards the development of
time resolved state-to-state measurements that will reveal the detailed flow of molecules
through state space as they relax in this quantum solvent.

\ack We acknowledge  Josef Tiggesb{\"a}umker and Carlo Callegari for carefully reading
the manuscript as well as Martina Havenith for helpful discussions. Furthermore, we thank
Marcel Drabbels, Joshua Jortner, Nadine Halberstadt, Martina Havenith, Alkwin Slenska,
Josef Tiggesb{\"a}umker and Andrej Vilesov for sharing unpublished work.\\

\bibliographystyle{fxunsrt}

\begin{thebibliography}{100}

\bibitem{Goyal:1992b}
S.~Goyal, D.~L. Schutt, and G.~Scoles.
\newblock {\em Phys.~Rev.~Lett.}, 69(6):933--936, 1992.

\bibitem{Whaley:1994}
K.~B. Whaley.
\newblock {\em International Reviews in Physical Chemistry}, 13:41, 1994.

\bibitem{Vilesov:1998b}
J.~P. Toennies and A.~F. Vilesov.
\newblock {\em Annu. Rev. Phys. Chem.}, 49:1--41, 1998.

\bibitem{Whaley:2000}
Yongkyung Kwon, Patrick Huang, Mehul~V. Patel, D\"{o}rte Blume, and K.~Birgitta
  Whaley.
\newblock {\em J. Chem. Phys.}, 113(16):6469 --6501, 2000.

\bibitem{Northby:2001}
J.~A. Northby.
\newblock {\em J.~Chem.~Phys.}, 115(22):10065--10077, 2001.

\bibitem{Callegari:2001}
C.~Callegari, K.~K. Lehmann, R.~Schmied, and G.~Scoles.
\newblock {\em J.~Chem.~Phys.}, 115(22):10090--10110, 2001.

\bibitem{Sti:2001e}
F.~Stienkemeier and A.~F. Vilesov.
\newblock {\em J. Chem. Phys.}, 115(17):10119--10137, 2001.

\bibitem{Ceperley:2001}
D.~M. Ceperley and E.~Manousakis.
\newblock {\em J.~Chem.~Phys.}, 115(22):10111--10118, 2001.

\bibitem{Dalfovo:2001}
F.~Dalfovo and S.~Stringari.
\newblock {\em J.~Chem.~Phys.}, 115(22):10078--10089, 2001.

\bibitem{Toennies:2001}
J.~P. Toennies, A.~F. Vilesov, and K.~B. Whaley.
\newblock {\em Physics Today}, 54(2):31--37, 2001.

\bibitem{Krotscheck:2001}
E.~Krotscheck and R.~Zillich.
\newblock {\em J.~Chem.~Phys.}, 115(22):10161--10174, 2001.

\bibitem{Toennies:2004}
J.~P. Toennies and A.~F. Vilesov.
\newblock {\em Angewandte Chemie}, 43(20):2622--2648, 2004.

\bibitem{Jortner:2005}
J.~Jortner and M.~Rosenblit.
\newblock {\em Advances in Chemical Physics}, in press, 2005.

\bibitem{Barranco:2006}
Manuel Barranco, Rafael Guardiola, Susana Hernandez, Ricardo Mayol, Jesus
  Navarro, and Marti Pi.
\newblock {\em J. Low Temp. Phys.}, in press, 2006.

\bibitem{Weilert:1996}
M.~A. Weilert, D.L. Whitaker, H.~J. Maris, and G.~M. Seidel.
\newblock {\em Phys.~Rev.~Lett.}, 77(23):4840–--4843, 1996.

\bibitem{Weilert:1995}
M.A. Weilert, D.~L. Whitaker, H.~Maris, and G.M. Seidel.
\newblock {\em J.~Low Temp.~Phys.}, 98, 1995.

\bibitem{Kim:2000}
H.~Kim, K.~Seo, B.~Tabbert, and G.~A. Williams.
\newblock {\em J.~Low Temp.~Phys.}, 121(5-6):621--626, 2000.

\bibitem{Kim:2002}
H.~Kim, K.~Seo, B.~Tabbert, and G.~A. Williams.
\newblock {\em Europhysics Letters}, 58(3):395--400, 2002.

\bibitem{Ghazarian:2002}
V.~Ghazarian, J.~Eloranta, and V.~A. Apkarian.
\newblock {\em Rev. Sci. Instr.}, 73:3606--3613, 2002.

\bibitem{Tsao:1998}
C.~C. Tsao, J.~D. Lobo, M.~Okumura, and S.~Y. Lo.
\newblock {\em J.~Phys.~D}, 31(17):2195--2204, 1998.

\bibitem{Boyle:1976}
F.P. Boyle and A.J. Dahm.
\newblock {\em Journal of Low Temperatures Physics}, 23:477, 1976.

\bibitem{Grisenti:2003}
R.~E. Grisenti and J.~P. Toennies.
\newblock {\em Phys.~Rev.~Lett.}, 90(23):34501, 2003.

\bibitem{Sti:2000b}
F.~Stienkemeier, M.~Wewer, F.~Meier, and H.~O. Lutz.
\newblock {\em Rev. Sci. Instr.}, 71:3480, 2000.

\bibitem{Harms:2001}
J.~Harms, J.~P. Toennies, M.~Barranco, and M.~Pi.
\newblock {\em Phys.~Rev.~B}, 63(18):184513, 2001.

\bibitem{Harms:1999a}
J.~Harms, M.~Hartmann, B.~Sartakov, J.~P. Toennies, and A.~F. Vilesov.
\newblock {\em J.~Chem.~Phys.}, 110(11):5124--5136, 1999.

\bibitem{Sti:2004d}
F.~Stienkemeier, O.~B{\"u}nermann, R.~Mayol, F.~Ancilotto, M.~Barranco, and
  M.~Pi.
\newblock {\em Phys. Rev. B}, 70:1, 2004.

\bibitem{Toennies:1990}
H.~Buchenau, E.~L. Knuth, J.~Northby, J.~P. Toennies, and C.~Winkler.
\newblock {\em J. Chem. Phys.}, 92:6875, 1990.

\bibitem{Harms:1996}
J.~Harms, J.~P. Toennies, and E.~L. Knuth.
\newblock {\em J.~Chem.~Phys.}, 106(8):3348--3357, 1996.

\bibitem{Harms:1998b}
J.~Harms, J.~P. Toennies, and F.~Dalfovo.
\newblock {\em Phys.~Rev.~B}, 58(6):3341--3350, 1998.

\bibitem{Lewerenz:1993}
M.~Lewerenz, B.~Schilling, and J.~P. Toennies.
\newblock {\em Chem.~Phys.~Lett.}, 206:381--287, 1993.

\bibitem{Knuth:1999}
E.~L. Knuth and U.~Henne.
\newblock {\em J.~Chem.~Phys.}, 110(5):2664--2668, 1999.

\bibitem{Slipchenko:2002}
M.~N. Slipchenko, S.~Kuma, T.~Momose, and A.~F. Vilesov.
\newblock {\em Rev.~Sci.~Instr.}, 73(10):3600--3605, 2002.

\bibitem{Gough:1985}
T.~E. Gough, M.~Mengel, P.~A. Rowntree, and G.~Scoles.
\newblock {\em J. Chem. Phys.}, 83(10):4958--4961, 1985.

\bibitem{Scheidemann:1990}
A.~Scheidemann, J.~P. Toennies, and J.~A. Northby.
\newblock {\em Phys. Rev. Lett.}, 64:1899, 1990.

\bibitem{Lewerenz:1995}
M.~Lewerenz, B.~Schilling, and J.~P. Toennies.
\newblock {\em J.~Chem.~Phys.}, 102(20):8191--8207, 1995.

\bibitem{Reho:2000a}
J.~H. Reho, U.~Merker, M.~R. Radcliff, K.~K. Lehmann, and G.~Scoles.
\newblock {\em J.~Phys.~Chem. A}, 104(16):3620--3626, 2000.

\bibitem{Kupper:2002}
J.~K{\"u}pper, J.~M. Merritt, and R.~E. Miller.
\newblock {\em J.~Chem.~Phys.}, 117(2):647--652, 2002.

\bibitem{Claas:2003}
P.~Claas, S.~O. Mende, and F.~Stienkemeier.
\newblock {\em Rev.~Sci.~Instr.}, 74(9):4071--4076, 2003.

\bibitem{Nauta:2000b}
K.~Nauta and R.~E. Miller.
\newblock {\em J.~Chem.~Phys.}, 113(22):10158--10168, 2000.

\bibitem{Sti:2003}
M.~Wewer and F.~Stienkemeier.
\newblock {\em Phys. Rev. B}, 67:125201, 2003.

\bibitem{Vongehr:2003}
S.~Vongehr and V.~V. Kresin.
\newblock {\em J.~Chem.~Phys.}, 119(21):11124--11129, 2003.

\bibitem{Diederich:2003}
Thomas Diederich.
\newblock PhD thesis, Universit{\"a}t Rostock, 2003.

\bibitem{Stringari:1987}
S.~Stringari and J.~Treiner.
\newblock {\em J. Chem. Phys.}, 87:5021, 1987.

\bibitem{Diederich:2001}
T.~Diederich, T.~D\"{o}ppner, J.~Braune, J.~Tiggesb\"{a}umker, and K.~H.
  Meiwes-Broer.
\newblock {\em Phys.~Rev.~Lett.}, 86(21):4807--4810, 2001.

\bibitem{Doppner:2001}
T.~D\"{o}ppner, T.~Diederich, J.~Tiggesb\"{a}umker, and K.~H. Meiwes-Broer.
\newblock {\em Eur.~Phys.~J.~D}, 16(1-3):13--16, 2001.

\bibitem{Diederich:2005}
T.~Diederich, T.~D\"{o}ppner, T.~Fennel, J.~Tiggesb\"{a}umker, and K.~H.
  Meiwes-Broer.
\newblock {\em Phys.~Rev.~A}, 72(2):11, August 2005.

\bibitem{Nauta:2001g}
K.~Nauta, D.~T. Moore, P.~L. Stiles, and R.~E. Miller.
\newblock {\em Science}, 292(5516):481--484, 2001.

\bibitem{Stiles:2003a}
P.~L. Stiles, D.~T. Moore, and R.~E. Miller.
\newblock {\em J.~Chem.~Phys.}, 118(17):7873--7881, 2003.

\bibitem{Nauta:2001f}
K.~Nauta and R.~E. Miller.
\newblock {\em J.~Chem.~Phys.}, 115(10):4508--4514, 2001.

\bibitem{Hartmann:1998}
M.~Hartmann, A.~Lindinger, J.~P. Toennies, and A.~F. Vilesov.
\newblock {\em Chem. Phys.}, 239(1-3):139--149, 1998.

\bibitem{Moore:2001a}
D.~T. Moore, M.~Ishiguro, and R.~E. Miller.
\newblock {\em J.~Chem.~Phys.}, 115(11):5144--5154, 2001.

\bibitem{Moore:2003b}
D.~T. Moore and R.~E. Miller.
\newblock {\em J.~Chem.~Phys.}, 119(9):4713--4721, 2003.

\bibitem{Diederich:2002}
T.~Diederich, J.~Tiggesb{\"a}umker, and K.~H. Meiwes-Broer.
\newblock {\em J.~Chem.~Phys.}, 116(8):3263--3269, 2002.

\bibitem{Mudrich:2004}
N.~Mudrich, O.~B{\"u}nermann, F.~Stienkemeier, O.~Dulieu, and
  M.~Weidem{\"u}ller.
\newblock {\em Eur.~Phys.~J.~D}, 31(2):291--299, 2004.

\bibitem{Moore:2003a}
D.~T. Moore and R.~E. Miller.
\newblock {\em J.~Chem.~Phys.}, 118(21):9629--9636, 2003.

\bibitem{Grebenev:2000d}
S.~Grebenev, B.~Sartakov, J.~P. Toennies, and A.~F. Vilesov.
\newblock {\em Science}, 289(5484):1532--1535, 2000.

\bibitem{Grebenev:2001b}
S.~Grebenev, B.~G. Sartakov, J.~P. Toennies, and A.~F. Vilesov.
\newblock {\em J.~Chem.~Phys.}, 114(2):617--620, 2001.

\bibitem{Grebenev:2002}
S.~Grebenev, B.~Sartakov, J.~P. Toennies, and A.~Vilesov.
\newblock {\em Phys.~Rev.~Lett.}, 89(22):225301, 2002.

\bibitem{Sti:1996a}
F.~Stienkemeier, J.~Higgins, C.~Callegari, S.I. Kanorsky, W.~E. Ernst, and
  G.~Scoles.
\newblock {\em Z. Phys. D.}, 38:253, 1996.

\bibitem{Bruhl:2001b}
F.~R. Br{\"u}hl, R.~A. Trasca, and W.~E. Ernst.
\newblock {\em J.~Chem.~Phys.}, 115(22):10220--10224, 2001.

\bibitem{Sti:1998a}
J.~Higgins, C.~Callegari, J.~Reho, F.~Stienkemeier, W.~E. Ernst, and G.~Scoles.
\newblock {\em J. Phys. Chem. A}, 102:4952, 1998.

\bibitem{Ancilotto:1995b}
F.~Ancilotto, E.~Cheng, M.~W. Cole, and F.~Toigo.
\newblock {\em Z. Phys. B}, 98:323, 1995.

\bibitem{Sti:1997b}
F.~Stienkemeier, F.~Meier, and H.~O. Lutz.
\newblock {\em J. Chem. Phys.}, 107(24):10816, 1997.

\bibitem{Sti:1999a}
F.~Stienkemeier, F.~Meier, and H.~O. Lutz.
\newblock {\em Eur. Phys. J. D}, 9:313, 1999.

\bibitem{Reho:2000d}
J.~Reho, U.~Merker, M.~R. Radcliff, K.~K. Lehmann, and G.~Scoles.
\newblock {\em J.~Chem.~Phys.}, 112(19):8409--8416, 2000.

\bibitem{Mella:2005}
M.~Mella, G.~Calderoni, and F.~Cargnoni.
\newblock {\em J.~Chem.~Phys.}, 123(5):054328, 2005.

\bibitem{Ancilotto:1995e}
F.~Ancilotto, P.~B. Lerner, and M.~W. Cole.
\newblock {\em J. Low Temp. Phys.}, 101:1123, 1995.

\bibitem{Harms:1997b}
J.~Harms, M.~Hartmann, J.~P. Toennies, A.~F. Vilesov, and B.~Sartakov.
\newblock {\em Journal Of Molecular Spectroscopy}, 185(1):204--206, 1997.

\bibitem{Guardiola:2000a}
R.~Guardiola and J.~Navarro.
\newblock {\em Phys.~Rev.~Lett.}, 84(6):1144--1147, 2000.

\bibitem{Barranco:1997a}
M.~Barranco, J.~Navarro, and A.~Poves.
\newblock {\em Phys.\ Rev.\ Lett.}, 78(25):4729--4732, 1997.

\bibitem{Guardiola:2000b}
R.~Guardiola.
\newblock {\em Phys.~Rev.~Lett.}, 62(5):3416--3421, August 2000.

\bibitem{Stienkemeier:2001}
F.~Stienkemeier and A.~F. Vilesov.
\newblock {\em J.~Chem.~Phys.}, 115(22):10119--10137, 2001.

\bibitem{vonHaeften:2001}
K.~von Haeften, T.~Laarmann, H.~Wabnitz, and T.~M{\"o}ller.
\newblock {\em Phys.~Rev.~Lett.}, 8715(15):153403, 2001.

\bibitem{vonHaeften:2002}
K.~von Haeften, T.~Laarmann, H.~Wabnitz, and T.~M{\"o}ller.
\newblock {\em Phys.~Rev.~Lett.}, 88(23):233401, 2002.

\bibitem{vonHaeften:2005}
K.~von Haeften, T.~Laarmann, H.~Wabnitz, and T.~M{\"o}ller.
\newblock {\em J.~Phys.~B}, 38(2):S373--S386, 2005.

\bibitem{Peterka:2003}
D.~S. Peterka, A.~Lindinger, L.~Poisson, M.~Ahmed, and D.~M. Neumark.
\newblock {\em Phys.~Rev.~Lett.}, 91(4):043401, 2003.

\bibitem{Nauta:1999c}
K.~Nauta and R.~E. Miller.
\newblock {\em Phys.~Rev.~Lett.}, 82(22):4480--4483, 1999.

\bibitem{Nauta:1999a}
K.~Nauta and R.~E. Miller.
\newblock {\em Science}, 283(5409):1895--1897, 1999.

\bibitem{Dong:2002}
F.~Dong and R.~E. Miller.
\newblock {\em Science}, 298(5596):1227--1230, 2002.

\bibitem{choi:2005a}
M.Y. Choi and R.E. Miller.
\newblock {\em Phys.~Chem.~Chem.~Phys.}, 7:3565--3573, 2005.

\bibitem{choi:2005b}
M.~Y. Choi, F.~Dong, and R.~E. Miller.
\newblock {\em Philosophical Transactions Of The Royal Society Of London Series
  A-Mathematical Physical And Engineering Sciences}, 363(1827):393--412, 2005.

\bibitem{Landau:1959}
L.~D. Landau and E.~M. Lifshitz.
\newblock {\em Statistical Physics}.
\newblock Pergamon Press, Ltd., 1959.

\bibitem{Brink:1990}
D.~M. Brink and S.~Stringari.
\newblock {\em Z.~Phys. D.}, 15:257--263, 1990.

\bibitem{Donnelly:1998}
R.~J. Donnelly and C.~F. Barenghi.
\newblock {\em Journal Physical Chemical Reference Data}, 27:1217--1274, 1998.

\bibitem{Barnett:1993a}
R.~N. Barnett and K.~B. Whaley.
\newblock {\em Phys.~Rev.~A}, 47:4082--4098, 1993.

\bibitem{Cheng:1996}
E.~Cheng, M.~A. McMahon, and K.~B. Whaley.
\newblock {\em J.~Chem.~Phys.}, 104:2669--2683, 1996.

\bibitem{Lamb:1916}
Horace Lamb.
\newblock {\em Hydrodynamics}.
\newblock Cambridge University Press, 1916.

\bibitem{Lehmann:2003a}
K.~K. Lehmann.
\newblock {\em J.~Chem.~Phys.}, 119(6):3336--3342, 2003.

\bibitem{Reatto:1999}
L.~Reatto and D.~E. Galli.
\newblock {\em Int. J. Mod. Phys. B}, 13(5\&6):607--616, 1999.

\bibitem{Hartmann:1995a}
M.~Hartmann, R.~E. Miller, J.~P. Toennies, and A.~F. Vilesov.
\newblock {\em Phys.~Rev.~Lett.}, 95:1566--1569, 1995.

\bibitem{Dalfovo:1994}
F.~Dalfovo.
\newblock {\em Z. Phys. D}, 29:61--66, 1994.

\bibitem{Kwon:1996}
Y.~K. Kwon, D.~M. Ceperley, and K.~B. Whaley.
\newblock {\em J.~Chem.~Phys.}, 104(6):2341--2348, 1996.

\bibitem{Lindinger:2001a}
A.~Lindinger, E.~Lugovoj, J.~P. Toennies, and A.~F. Vilesov.
\newblock {\em Z. Phys. Chem.}, 215:401--416, 2001.

\bibitem{Huang:2004}
P.~Huang, H.~D. Whitley, and K.~B. Whaley.
\newblock {\em J.~Low Temp.~Phys.}, 134(1-2):263--268, 2004.

\bibitem{Blume:1999}
D.~Blume, M.~Mladenovic, M.~Lewerenz, and K.~B. Whaley.
\newblock {\em J.~Chem.~Phys.}, 110(12):5789--5805, 1999.

\bibitem{Schmied:2005}
R. Schmied and K.K. Lehmann, work in progress.

\bibitem{Chin:1995a}
S.~A. Chin and E.~Krotscheck.
\newblock {\em Phys.~Rev.~Lett.}, 74(7):1143--1146, 1995.

\bibitem{Chin:1995b}
S.~A. Chin and E.~Krotscheck.
\newblock {\em Phys.~Rev.~B}, 52(14):10405--10428, 1995.

\bibitem{Wyatt:1992}
A.~F.~G. Wyatt.
\newblock {\em J.~Low Temp.~Phys.}, 87:453--472, 1992.

\bibitem{Hartmann:1999}
M.~Hartmann, N.~P{\"o}rtner, B.~Sartakov, J.~P. Toennies, and A.~F. Vilesov.
\newblock {\em J.~Chem.~Phys.}, 110(11):5109--5123, 1999.

\bibitem{Lehmann:2004b}
K.~K. Lehmann and A.~M. Dokter.
\newblock {\em Phys.~Rev.~Lett.}, 92(17):173401, 2004.

\bibitem{Bauer:1979}
A.~Bauer, D.~Boucher, J.~Burie, J.~Demaison, A.~Dubrulle, G.~H. Bauer, R.~J.
  Donnelly, and W.~F. Vien.
\newblock {\em J. Phys.\ Chem.\ Ref.\ Data}, 8:537, 1979.

\bibitem{Pi:2000}
M.~Pi, R.~Mayol, M.~Barranco, and F.~Dalfovo.
\newblock {\em J.~Low Temp.~Phys.}, 121:423--428, 2000.

\bibitem{Lehmann:2003b}
K.~K. Lehmann and R.~Schmied.
\newblock {\em Phys.~Rev.~B}, 68(22):224520, 2003.

\bibitem{Rayfield:1964}
G.~W. Rayfield and F.~Reif.
\newblock {\em Phys.~Rev.~A}, 136:1194--1208, 1964.

\bibitem{Dalfovo:2000}
F.~Dalfovo, R.~Mayol, M.~Pi, and M.~Barranco.
\newblock {\em Phys.~Rev.~Lett.}, 85(5):1028--1031, 2000.

\bibitem{Ancilotto:2003}
F.~Ancilotto, M.~Barranco, and M.~Pi.
\newblock {\em Phys.~Rev.~Lett.}, 91(10):105302, 2003.

\bibitem{Lehmann:1999}
K.~K. Lehmann.
\newblock {\em Mol. Phys.}, 97(5):645--666, 1999.

\bibitem{Lehmann:2000}
K.~K. Lehmann.
\newblock {\em Mol. Phys.}, 98(23):1991--1993, 2000.

\bibitem{Toennies:1995}
J.~P. Toennies and A.~F. Vilesov.
\newblock {\em Chem.~Phys.~Lett.}, 235(5-6):596--603, 1995.

\bibitem{Dicke:1953}
R.H. Dicke.
\newblock {\em Phys. Rev.}, 89:47, 1953.

\bibitem{Stienkemeier:1995c}
F.~Stienkemeier, J.~Higgins, W.~E. Ernst, and G.~Scoles.
\newblock {\em Z. Phys. B}, 98(3):413--416, 1995.

\bibitem{Hartmann:1996a}
M.~Hartmann, F.~Mielke, J.~P. Toennies, A.~F. Vilesov, and G.~Benedek.
\newblock {\em Phys.~Rev.~Lett.}, 76(24):4560--4563, 1996.

\bibitem{Baer:1996}
Tomas Baer and William~L. Hase.
\newblock {\em Unimolecular reaction dynamics : theory and experiments}.
\newblock Oxford University Press, New York, 1996.

\bibitem{REMiller_pc}
Roger~E. Miller.
\newblock private communication, 2005.

\bibitem{Whaley:1998}
K.~B. Whaley.
\newblock {\em Advances in Molecular Vibrations and Collision Dynamics},
  III:397--451, 1998.

\bibitem{Portner:2003}
N.~P{\"o}rtner, A.~F. Vilesov, and M.~Havenith.
\newblock {\em Chem.~Phys.~Lett.}, 368(3-4):458--464, 2003.

\bibitem{Lehmann:2004a}
K.~K. Lehmann.
\newblock {\em J.~Chem.~Phys.}, 120(2):513--515, 2004.

\bibitem{Tisza:1938}
L.~Tisza.
\newblock {\em Nature}, 141:913, 1938.

\bibitem{Andronikashvili:1946}
E.~L. Andronikashvili.
\newblock {\em J. Phys. USSR}, 10:201, 1946.

\bibitem{Ceperley:1995}
D.~M. Ceperley.
\newblock {\em Rev. Mod. Phys.}, 67:279--355, 1995.

\bibitem{Feynman:1990}
R.~P. Feynman.
\newblock {\em Statistical Mechanics: A Set of Lectures}.
\newblock 1990.

\bibitem{Sindzingre:1989}
P.~Sindzingre, M.~L. Klein, and D.~M. Ceperley.
\newblock {\em Phys.~Rev.~Lett.}, 63:1601--1604, 1989.

\bibitem{Grebenev:1998}
S.~Grebenev, J.~P. Toennies, and A.~F. Vilesov.
\newblock {\em Science}, 279(5359):2083--2086, 1998.

\bibitem{Kwon:1999b}
Y.~Kwon and K.~B. Whaley.
\newblock {\em Phys.~Rev.~Lett.}, 83:4108--4111, 1999.

\bibitem{Kwon:2000}
Y.~Kwon, P.~Huang, M.~V. Patel, D.~Blume, and K.~B. Whaley.
\newblock {\em J.~Chem.~Phys.}, 113:6469--6501, 2000.

\bibitem{Kwon:2001b}
Y.~Kwon and K.~B. Whaley.
\newblock {\em J.~Chem.~Phys.}, 115:10146--10153, 2001.

\bibitem{Kwon:2003}
Y.~Kwon and K.~B. Whaley.
\newblock {\em J.~Chem.~Phys.}, 119(4):1986--1995, 2003.

\bibitem{Draeger:2003}
E.~W. Draeger and D.~M. Ceperley.
\newblock {\em Phys.~Rev.~Lett.}, 90(6), 2003.

\bibitem{Miller:2001}
R.~E. Miller.
\newblock {\em Faraday Discussions}, 118:1--17, 2001.

\bibitem{Draeger:up}
E.W. Draeger, K.K. Lehmann, and R.E. Miller.
\newblock The microscopic {A}ndronikashvili experiment as a probe of superfluid
  excitations.
\newblock Unpublished.

\bibitem{Kwon:2005}
Y.~Kwon and K.~B. Whaley.
\newblock {\em J.~Low Temp.~Phys.}, 138(1-2):253--258, 2005.

\bibitem{Callegari:1999}
C.~Callegari, A.~Conjusteau, I.~Reinhard, K.~K. Lehmann, G.~Scoles, and
  F.~Dalfovo.
\newblock {\em Phys.~Rev.~Lett.}, 83:5058--5061, 1999.

\bibitem{Callegari:2000a}
C.~Callegari, A.~Conjusteau, I.~Reinhard, K.~K. Lehmann, G.~Scoles, and
  F.~Dalfovo.
\newblock {\em Phys.~Rev.~Lett.}, 84:1848, 2000.

\bibitem{Dalfovo:1995}
F.~Dalfovo, A.~Lastri, L.~Pricaupenko, S.~Stringari, and J.~Treiner.
\newblock {\em Phys.~Rev.~B}, 52(2):1193--1209, 1995.

\bibitem{Conjusteau:2000}
A.~Conjusteau, C.~Callegari, I.~Reinhard, K.~K. Lehmann, and G.~Scoles.
\newblock {\em J.~Chem.~Phys.}, 113(12):4840--4843, 2000.

\bibitem{Viel:2001}
A.~Viel and K.~B. Whaley.
\newblock {\em J.~Chem.~Phys.}, 115:10186--10198, 2001.

\bibitem{Lehmann:2001c}
K.~K. Lehmann.
\newblock {\em Faraday Discussions}, 118:33--42, 2001.

\bibitem{Rossi:2004}
M.~Rossi, M.~Verona, D.~E. Galli, and L.~Reatto.
\newblock {\em Phys.~Rev.~B}, 69(21):212510, 2004.

\bibitem{Lehmann:2002}
K.~K. Lehmann and C.~Callegari.
\newblock {\em J.~Chem.~Phys.}, 117(4):1595--1603, 2002.

\bibitem{Paesani:2004b}
F.~Paesani and K.~B. Whaley.
\newblock {\em J.~Chem.~Phys.}, 121(11):5293--5311, 2004.

\bibitem{Paesani:2005a}
F.~Paesani, Y.~Kwon, and K.~B. Whaley.
\newblock {\em Phys.~Rev.~Lett.}, 94(15):153401, 2005.

\bibitem{Paesani:2005b}
F.~Paesani, R.~E. Zillich, Y.~Kwon, and K.~B. Whaley.
\newblock {\em J.~Chem.~Phys.}, 122(18):181106, 2005.

\bibitem{Moroni:2003}
S.~Moroni, A.~Sarsa, S.~Fantoni, K.~E. Schmidt, and S.~Baroni.
\newblock {\em Phys. Rev. Lett.}, 90:143401, 2003.

\bibitem{Moroni:2004}
S.~Moroni, N.~Blinov, and P.-N. Roy.
\newblock {\em J. Chem. Phys.}, 121:3577--3581, 2004.

\bibitem{Cazzato:2004}
P.~Cazzato, S.~Paolini, S.~Moroni, and S.~Baroni.
\newblock {\em J. Chem. Phys.}, 120:9071--9076, 2004.

\bibitem{Paolini:2005}
S.~Paolini, S.~Fantoni, S.~Moroni, and S.~Baroni.
\newblock {\em J. Chem. Phys.}, 123:114306, 2005.

\bibitem{Moroni:2005}
S.~Moroni and S.~Baroni.
\newblock {\em Computer Physics Communications}, 169:404--407, 2005.

\bibitem{Blinov:2004}
N.~Blinov, X.~Song, and P.-N. Roy.
\newblock {\em J. Chem. Phys.}, 120:5916, 2004.

\bibitem{Blinov:2005}
N.~Blinov and P.-N. Roy.
\newblock {\em J. Low Temp. Phys.}, 140:253--267, 2005.

\bibitem{Zillich:2004b}
R.~E. Zillich, Y.~Kwon, and K.~B. Whaley.
\newblock {\em Phys. Rev. Lett.}, 93:250401, 2004.

\bibitem{Zillich:2005}
R.~E. Zillich, F.~Paesani, Y.~Kwon, and K.~B. Whaley.
\newblock {\em J. Chem. Phys.}, 123:114301, 2005.

\bibitem{Zillich:2004a}
R.~E. Zillich and K.~B. Whaley.
\newblock {\em Phys. Rev. B}, page 104517, 2004.

\bibitem{Sti:unp_fsK2}
P.~Claas, C.~P. Schulz, and F.~Stienkemeier.
\newblock {\em Wavepacket dynamics of {K$_2$} attached to helium nanodroplets},
  unpublished results.

\bibitem{Schreiber:1997}
S.~Rutz and E.~Schreiber.
\newblock {\em Chem. Phys. Lett.}, 269:9--16, 1997.

\bibitem{Sti:2004}
C.~P. Schulz, P.~Claas, D.~Schumacher, and F.~Stienkemeier.
\newblock {\em Phys. Rev. Lett.}, 92(1):013401, 2004.

\bibitem{Sti:unp_fsNa2}
P.~Claas, C.~P. Schulz, and F.~Stienkemeier.
\newblock {\em Wavepacket dynamics of {Na$_2$} attached to helium
  nanodroplets}, unpublished results.

\bibitem{Kuehling:1993}
H.~K{\"u}hling, S.~Rutz, K.~Kobe, E.~Schreiber, and L.~W{\"o}ste.
\newblock {\em J.~Phys.~Chem.}, 97(48):12500--12503, 1993.

\bibitem{Sti:unp_fsKn}
P.~Claas, C.~P. Schulz, and F.~Stienkemeier.
\newblock {\em Fragmentation dynamics of potassium clusters attached to helium
  droplets}, unpublished results.

\bibitem{Kuehling:1994}
H.~K{\"u}hling, K.~Kobe, S.~Rutz, E.~Schreiber, and L.~W{\"o}ste.
\newblock {\em J.~Phys.~Chem.}, 98(27):6679, 1994.

\bibitem{Conjusteau_thesis}
Andre Conjusteau.
\newblock {\em Spectroscopy in Helium Nanodroplets: Studying Relaxation
  Mechanisms in Nature's Most Fascinating Solvent}.
\newblock PhD thesis, Princeton University, 2002.

\bibitem{Stolyarov:2004}
D.~Stolyarov, E.~Polyakova, and C.~Wittig.
\newblock {\em J.~Phys.~Chem. A}, 108(45):9841--9846, 2004.

\bibitem{Sti:1999b}
F.~Stienkemeier, F.~Meier, A.~H{\"a}gele, H.~O. Lutz, E.~Schreiber, C.~P.
  Schulz, and I.~V. Hertel.
\newblock {\em Phys. Rev. Lett.}, 83(12):2320, 1999.

\bibitem{Sti:unp_fsCs}
G.~Droppelmann, C.~P. Schulz, and F.~Stienkemeier.
\newblock {\em Surface dynamics of helium nanodroplets upon femtosecond
  excitation of attached alkali atoms}, unpublished results.

\bibitem{Sti:2004b}
G.~Droppelmann, O.~B{\"u}nermann, C.~P. Schulz, and F.~Stienkemeier.
\newblock {\em Phys. Rev. Lett.}, 93:023402, 2004.

\bibitem{Sti:1998b}
C.~Callegari, J.~Higgins, F.~Stienkemeier, and G.~Scoles.
\newblock {\em J. Phys. Chem. A}, 102:95, 1998.

\bibitem{Lehmann:2000c}
J.~Reho, J.~Higgins, C.~Callegari, K.~K. Lehmann, and G.~Scoles.
\newblock {\em J. Chem. Phys.}, 113(21):9686--9693, 2000.

\bibitem{Sti:2001d}
C.~P. Schulz, P.~Claas, and F.~Stienkemeier.
\newblock {\em Phys. Rev. Lett.}, 87(15):153401, 2001.

\bibitem{Nettels:2005}
D.~Nettels, A.~Hofer, P.~Moroshkin, R.~M{\"u}ller-Siebert, S.~Ulzega, and
  A.~Weis.
\newblock {\em Phys.~Rev.~Lett.}, 94(6):063001, 2005.

\bibitem{Lehmann:1997}
J.~Reho, C.~Callegari, J.~Higgins, W.~E. Ernst, K.~K. Lehmann, and G.~Scoles.
\newblock {\em Faraday Discuss.}, 108:161--174, 1997.

\bibitem{Lehmann:2000d}
J.~Reho, J.~Higgins, C.~Callegari, K.~K. Lehmann, and G.~Scoles.
\newblock {\em J. Chem. Phys.}, 113(21):9694--9701, 2000.

\bibitem{Sti:2000d}
J.~Higgins, J.~Reho, C.~Callegari, F.~Stienkemeier, W.E. Ernst, K.K. Lehmann,
  and G.~Scoles.
\newblock {\em Spectroscopy in, on, and off a Beam of Superfluid Helium
  Nanodroplets}, pages 723--754.
\newblock in: Atomic and Molecular Beams, The State of the Art 2000, R.
  Campargue (ed.), Springer, 2000.

\bibitem{Sti:unp_qi}
P.~Claas, G.~Droppelmann, M.~Mudrich, C.P. Schulz, and F.~Stienkemeier.
\newblock {\em Quantum interference oscillations revealing the vibrational
  structure of unstable {RbHe} exciplex molecules}, unpublished results.

\bibitem{Bonhommeau:2004}
D.~Bonhommeau, A.~Viel, and N.~Halberstadt.
\newblock {\em J.~Chem.~Phys.}, 120(24):11359--11362, 2004.

\bibitem{Ruchti:1998}
T.~Ruchti, K.~Forde, B.~E. Callicoatt, H.~Ludwigs, and K.~C. Janda.
\newblock {\em J.~Chem.~Phys.}, 109(24):10679--10687, 1998.

\bibitem{Reho:2001}
J.~H. Reho, J.~P. Higgins, and K.~K. Lehmann.
\newblock {\em Faraday Discussions}, 118:33--42, 2001.

\bibitem{Reho:2001a}
J.~H. Reho, J.~P. Higgins, M.~Nooijen, K.~K. Lehmann, G.~Scoles, and
  M.~Gutowski.
\newblock {\em J. Chem. Phys.}, 115:10265--10274, 2001.

\bibitem{Braun:2004}
A.~Braun and M.~Drabbels.
\newblock {\em Phys.~Rev.~Lett.}, 93(25), 2004.

\bibitem{Braun:2004a}
A.~Braun.
\newblock {\em Photodissociation studies of {CH$_3$I} and {CF$_3$I} in fluid
  $^4${H}elium nanodroplets}.
\newblock PhD thesis, EPFL Lausanne, 2004.

\bibitem{Alexander:1988}
M.~L. Alexander, N.~E. Levinger, M.~A. Johnson, D.~Ray, and W.~C. Lineberger.
\newblock {\em J.~Chem.~Phys.}, 88(10):6200--6210, 1988.

\bibitem{Buck:2002}
U.~Buck.
\newblock {\em J.~Phys.~Chem.}, 106(43):10049--10062, 2002.

\bibitem{Vorsa:1996}
V.~Vorsa, P.~J. Campagnola, S.~Nandi, M.~Larsson, and W.~C. Lineberger.
\newblock {\em J.~Chem.~Phys.}, 105(6):2298--2308, 1996.

\bibitem{Radcliffe:2004}
P.~Radcliffe, A.~Przystawik, T.~Diederich, T.~D{\"o}ppner,
  J.~Tiggesb{\"a}umker, and K.~H. Meiwes-Broer.
\newblock {\em Phys.~Rev.~Lett.}, 92(17):173403, 2004.

\bibitem{Przystawik:2005}
A.~Przystawik, P.~Radcliffe, T.~Diederich, T.~D{\"o}ppner, Josef
  Tiggesb{\"a}umker, and K.-H. Meiwes-Broer.
\newblock Lifetime dynamics of the {Ag$_3$} excited state manifold.
\newblock unpublished results, 2005.

\bibitem{Loginov:2005}
E.~Loginov, D.~Rossi, and M.~Drabbels.
\newblock Photoelectron spectroscopy of doped helium nanodroplets.
\newblock unpublished results.

\bibitem{Lehmann:1999b}
K.~K. Lehmann and J.~A. Northby.
\newblock {\em Mol. Phys.}, 97(5):639--644, 1999.

\bibitem{Farnik:1999}
M.~Farnik, B.~Samelin, and J.~P. Toennies.
\newblock {\em J.~Chem.~Phys.}, 110(18):9195--9201, 1999.

\bibitem{Farnik:2003}
M.~Farnik and J.~P. Toennies.
\newblock {\em J.~Chem.~Phys.}, 118(9):4176--4182, 2003.

\bibitem{Rosenbilt:1994b}
M.~Rosenbilt and Jortner Joshua.
\newblock {\em J.~Chem.~Phys.}, 101:9982--9996, 1994.

\bibitem{Lippert:2003}
H.~Lippert, V.~Stert, L.~Hesse, C.~P. Schulz, I.~V. Hertel, and W.~Radloff.
\newblock {\em Chem. Phys. Lett.}, 371(1-2):208--216, 2003.

\bibitem{Sobolewski:2002}
A.~L. Sobolewski, W.~Domcke, C.~Dedonder-Lardeux, and C.~Jouvet.
\newblock {\em Phys. Chem. Chem. Phys.}, 4(7):1093--1100, 2002.

\bibitem{Nibbering:2000}
E.~T.~J. Nibbering, F.~Tschirschwitz, C.~Chudoba, and T.~Elsaesser.
\newblock {\em J. Phys. Chem. A}, 104(18):4236--4246, 2000.

\bibitem{Kim:1995a}
S.~K. Kim, J.~J. Breen, D.~M. Willberg, L.~W. Peng, A.~Heikal, J.~A. Syage, and
  A.~H. Zewail.
\newblock {\em J. Phys. Chem.}, 99(19):7421--7435, 1995.

\bibitem{Lindinger:1999}
A.~Lindinger, J.~P. Toennies, and A.~F. Vilesov.
\newblock {\em J.~Chem.~Phys.}, 110(3):1429--1436, 1999.

\bibitem{Sti:2004a}
M.~Wewer and F.~Stienkemeier.
\newblock {\em J. Chem. Phys.}, 120(3):1239--1244, 2004.

\bibitem{Lehnig:2004c}
R.~Lehnig, M.~Slipchenko, S.~Kuma, T.~Momose, B.~Sartakov, and A.F. Vilesov.
\newblock {\em J.~Chem.~Phys.}, 121(19):9396--9405, 2004.

\bibitem{Sti:2005}
F.~Stienkemeier and M.~Wewer.
\newblock {\em Phys. Chem. Chem. Phys.}, 7(6):1171--1175, 2005.

\bibitem{Lehnig:2003}
R.~Lehnig and A.~Slenczka.
\newblock {\em J.~Chem.~Phys.}, 118(18):8256--8260, 2003.

\bibitem{Lehnig:2004a}
R.~Lehnig and A.~Slenczka.
\newblock {\em Chem. Phys. Chem.}, 5(7):1014--1019, 2004.

\bibitem{Lehnig:2005}
R.~Lehnig and A.~Slenczka.
\newblock {\em J.~Chem.~Phys.}, 122:244317, 2005.

\bibitem{Whitley:2005}
H.~D. Whitley, P.~Huang, Y.~Kwon, and K.~B. Whaley.
\newblock {\em J. Chem. Phys.}, 123:54307, 2005.

\bibitem{Hartmann:2002}
M.~Hartmann, A.~Lindinger, J.~P. Toennies, and A.~F. Vilesov.
\newblock {\em Phys.~Chem.~Chem.~Phys.}, 4(20):4839--4844, 2002.

\bibitem{Hartmann:2001}
M.~Hartmann, A.~Lindinger, J.~P. Toennies, and A.~F. Vilesov.
\newblock {\em J.~Phys.~Chem. A}, 105(26):6369--6377, 2001.

\bibitem{Lindinger:2001b}
A.~Lindinger, J.~P. Toennies, and A.~F. Vilesov.
\newblock {\em Phys.~Chem.~Chem.~Phys.}, 3(13):2581--2587, 2001.

\bibitem{Portner:2002}
N.~P{\"o}rtner, J.~P. Toennies, and A.~F. Vilesov.
\newblock {\em J.~Chem.~Phys.}, 117(13):6054--6060, 2002.

\bibitem{Slenczka:2001}
A.~Slenczka, B.~Dick, M.~Hartmann, and J.~P. Toennies.
\newblock {\em J.~Chem.~Phys.}, 115(22):10199--10205, 2001.

\bibitem{Dick:2001}
B.~Dick and A.~Slenczka.
\newblock {\em J.~Chem.~Phys.}, 115(22):10206--10213, 2001.

\bibitem{Lindinger:2004}
A.~Lindinger, J.~P. Toennies, and A.~F. Vilesov.
\newblock {\em J.~Chem.~Phys.}, 121(24):12282--12292, 2004.

\bibitem{Carcabal:2004}
P.~Carcabal, R.~Schmied, K.~K. Lehmann, and G.~Scoles.
\newblock {\em J.~Chem.~Phys.}, 120(14):6792--6793, 2004.

\bibitem{Schmied:2004}
R.~Schmied, P.~Carcabal, A.~M. Dokter, V.~P.~A. Lonij, K.~K. Lehmann, and
  G.~Scoles.
\newblock {\em J.~Chem.~Phys.}, 121(6):2701--2710, 2004.

\bibitem{Boatwright:2005}
A.~Boatwright, N.~A. Besley, S.~Curtis, R.~R. Wright, and A.~J. Stace.
\newblock {\em J. Chem. Phys.}, 123:21102, 2005.

\bibitem{Curtis:2005}
S.~Curtis, A.~Boatwright, R.~R. Wright, and A.~J. Stace.
\newblock {\em Chem. Phys. Lett.}, 401(1-3):254--258, 2005.

\bibitem{Krasnokutski:2005}
S.~Krasnokutski, G.~Rouille, and F.~Huisken.
\newblock {\em Chem.~Phys.~Lett.}, 406(4-6):386--392, 2005.

\bibitem{Kanya:2004}
R.~Kanya and Y.~Ohshima.
\newblock {\em J.~Chem.~Phys.}, 121(19):9489--9497, 2004.

\bibitem{Lehnig:2004}
R.~Lehnig and A.~Slenczka.
\newblock Unpublished work.

\bibitem{Callicoatt:1996}
B.~E. Callicoatt, D.~D. Mar, V.~A. Apkarian, and K.~C. Janda.
\newblock {\em J.~Chem.~Phys.}, 105:7872--7875, 1996.

\bibitem{Callicoatt:1998a}
B.~E. Callicoatt, K.~Forde, L.~F. Jung, T.~Ruchti, and K.~C. Janda.
\newblock {\em J.~Chem.~Phys.}, 109(23):10195--10200, 1998.

\bibitem{Lewis:2005}
W.~K. Lewis, C.~M. Lindsay, R.~J. Bemish, and R.~E. Miller.
\newblock {\em Journal Of The American Chemical Society}, 127(19):7235--7242,
  2005.

\bibitem{Lewis:2005a}
W.~W.K.~Lewis, R.J. Bemish, and R.E. Miller.
\newblock {\em J. Chem. Phys.}, 123:141103, 2005.

\bibitem{Lewis:2004}
W.~K. Lewis, B.~E. Applegate, J.~Sztaray, B.~Sztaray, T.~Baer, R.~J. Bemish,
  and R.~E. Miller.
\newblock {\em Journal Of The American Chemical Society}, 126(36):11283--11292,
  2004.

\bibitem{Merritt:2004}
J.~M. Merritt, G.~E. Douberly, and R.~E. Miller.
\newblock {\em J.~Chem.~Phys.}, 121(3):1309--1316, 2004.

\bibitem{Douberly:2005a}
G.~E. Douberly, J.~M. Merritt, and R.~E. Miller.
\newblock {\em Phys. Chem. Chem. Phys.}, 7(3):463--468, 2005.

\bibitem{Douberly:2005b}
G.~E. Douberly and R.~E. Miller.
\newblock {\em J.~Chem.~Phys.}, 122(2), 2005.

\bibitem{Nauta:1999}
K.~Nauta, D.~T. Moore, and R.~E. Miller.
\newblock {\em Faraday Discussions}, 113:261--278, 1999.

\bibitem{Madeja:2004}
F.~Madeja, M.~Havenith, K.~Nauta, R.~E. Miller, J.~Chocholousova, and P.~Hobza.
\newblock {\em J.~Chem.~Phys.}, 120(22):10554--10560, 2004.

\bibitem{Douberly:2003a}
G.~E. Douberly and R.~E. Miller.
\newblock {\em J.~Phys.~Chem. B}, 107(19):4500--4507, 2003.

\bibitem{Nauta:2000a}
K.~Nauta and R.~E. Miller.
\newblock {\em Science}, 287(5451):293--295, 2000.

\bibitem{Burnham:2002}
C.~J. Burnham, S.~S. Xantheas, M.~A. Miller, B.~E. Applegate, and R.~E. Miller.
\newblock {\em J.~Chem.~Phys.}, 117(3):1109--1122, 2002.

\bibitem{Merritt:2005}
J.~M. Merritt, J.~K{\"u}pper, and R.~E. Miller.
\newblock {\em Phys.~Chem.~Chem.~Phys.}, 7(1):67--78, 2005.

\bibitem{Stiles:2004}
P.~L. Stiles, D.~T. Moore, and R.~E. Miller.
\newblock {\em J.~Chem.~Phys.}, 121(7):3130--3142, 2004.

\bibitem{Dong:2004b}
F.~Dong and R.~E. Miller.
\newblock {\em J.~Phys.~Chem. A}, 108(12):2181--2191, 2004.

\bibitem{Moore:2004b}
D.~T. Moore and R.~E. Miller.
\newblock {\em J.~Phys.~Chem. A}, 108(45):9908--9915, 2004.

\bibitem{Grebenev:2001a}
S.~Grebenev, E.~Lugovoi, B.~G. Sartakov, J.~P. Toennies, and A.~F. Vilesov.
\newblock {\em Faraday Discussions}, 118:19--32, 2001.

\bibitem{Grebenev:2003}
S.~Grebenev, B.~G. Sartakov, J.~P. Toennies, and A.~F. Vilesov.
\newblock {\em J.~Chem.~Phys.}, 118(19):8656--8670, 2003.

\bibitem{Paesani:2003b}
F.~Paesani, R.~E. Zillich, and K.~B. Whaley.
\newblock {\em J.~Chem.~Phys.}, 119(22):11682--11694, 2003.

\bibitem{Moore:2001b}
D.~T. Moore, M.~Ishiguro, L.~Oudejans, and R.~E. Miller.
\newblock {\em J.~Chem.~Phys.}, 115:5137--5143, 2001.

\bibitem{Moore:2004a}
D.~T. Moore and R.~E. Miller.
\newblock {\em J.~Phys.~Chem. A}, 108(11):1930--1937, 2004.

\bibitem{Madeja:2002}
F.~Madeja, P.~Markwick, M.~Havenith, K.~Nauta, and R.~E. Miller.
\newblock {\em J.~Chem.~Phys.}, 116(7):2870--2878, 2002.

\bibitem{Lindsay:2005}
C.~M. Lindsay and R.~E. Miller.
\newblock {\em J.~Chem.~Phys.}, 122(10):104306, 2005.

\bibitem{Scheele:2005}
I.~Scheele, A.~Conjusteau, C.~Callegari, R.~Schmied, K.~K. Lehmann, and
  G.~Scoles.
\newblock {\em J.~Chem.~Phys.}, 122(10):104307, 2005.

\bibitem{Slipchenko:2005}
M.~N. Slipchenko and A.~F. Vilesov.
\newblock {\em Chem. Phys. Lett.}, 412:176--183, 2005.

\bibitem{Behrens:1998}
M.~Behrens, U.~Buck, R.~Fr{\"o}chtenicht, M.~Hartmann, F.~Huisken, and
  F.~Rohmund.
\newblock {\em J.~Chem.~Phys.}, 110:5914--5920, 1998.

\bibitem{vonHaeften:2005b}
K.~von Haeften, A.~Merzelthin, S.~Rudolph, V.~Staemmler, and M.~Havenith.
\newblock {\em Phys. Rev. Lett.}, 95:21501, 2005.

\bibitem{vonHaeften:2005c}
K.~von Haeften, S.~Rudolph, I.~Simanovki, M.~Havenith, R.E. Zillich, and K.B.
  Whaley.
\newblock {\em Phys. Rev. B}, in press, 2006.

\bibitem{Paesani:2002}
F.~Paesani and F.~A. Gianturco.
\newblock {\em J.~Chem.~Phys.}, 116(23):10170--10182, 2002.

\bibitem{Tang:2003a}
J.~Tang and A.~R.~W. McKellar.
\newblock {\em J.~Chem.~Phys.}, 119(2):754--764, 2003.

\bibitem{Rabitz:1970}
H.A. Rabitz and R.G. Gordon.
\newblock {\em J. Chem. Phys.}, 53:1831--1850, 1970.

\bibitem{Nauta:2004}
K.~Nauta and R.~E. Miller.
\newblock {\em Journal Of Molecular Spectroscopy}, 223(1):101--105, 2004.

\bibitem{Stiles:2003b}
P.~L. Stiles, K.~Nauta, and R.~E. Miller.
\newblock {\em Phys.~Rev.~Lett.}, 90(13):135301, 2003.

\bibitem{Kuyanov:up}
K.E. Kuyanov, M.N. Slipchenko, and A.F. Vilesov.
\newblock Spectra of the $\nu_1$ and $\nu_3$ bands of water molecules in helium
  droplets.
\newblock preprint, 2006.

\bibitem{Hoshina:2005}
H.~Hoshina, J.~Lucrezi, M.~N. Slipchenko, K.~E. Kuyanov, and A.~F. Vilesov.
\newblock {\em Phys.~Rev.~Lett.}, 94(19):195301, 2005.

\bibitem{Lehmann:2001}
K.~K. Lehmann.
\newblock {\em J.~Chem.~Phys.}, 114(10):4643--4648, 2001.

\bibitem{Schollkopf:1996}
W.~Schollkopf and J.~P. Toennies.
\newblock {\em J.~Chem.~Phys.}, 104(3):1155--1158, 1996.

\bibitem{Bruhl:2004}
R.~Br{\"u}hl, R.~Guardiola, A.~Kalinin, O.~Kornilov, J.~Navarro, T.~Savas, and
  J.~P. Toennies.
\newblock {\em Phys.~Rev.~Lett.}, 92(18):~, 2004.

\bibitem{Sola:2004}
E.~Sola, J.~Casulleras, and J.~Boronat.
\newblock {\em J.~Low Temp.~Phys.}, 134(1-2):787--792, 2004.

\bibitem{Tang:2002d}
J.~Tang, Y.~J. Xu, A.~R.~W. McKellar, and W.~J{\"a}ger.
\newblock {\em Science}, 297(5589):2030--2033, 2002.

\bibitem{Tang:2003b}
J.~Tang and A.~R.~W. McKellar.
\newblock {\em J.~Chem.~Phys.}, 119(11):5467--5477, 2003.

\bibitem{Xu:2003}
Y.~J. Xu and W.~J{\"a}ger.
\newblock {\em J.~Chem.~Phys.}, 119(11):5457--5466, 2003.

\bibitem{Xu:2003b}
Y.J. Xu, W.~J{\"a}ger, J.~Tang, and A.R.W. McKellar.
\newblock {\em Phys. Rev. Lett.}, 91:163401, 2003.

\bibitem{Tang:2004a}
J.~Tang and A.~R.~W. McKellar.
\newblock {\em J.~Chem.~Phys.}, 121(1):181--190, 2004.

\bibitem{Tang:2004c}
J.~Tang, A.~R.~W. McKellar, E.~Mezzacapo, and S.~Moroni.
\newblock {\em Phys.~Rev.~Lett.}, 92(14):145503, 2004.

\bibitem{McKellar:2004}
A.~R.~W. McKellar.
\newblock {\em J.~Chem.~Phys.}, 121(14):6868--6873, 2004.

\end{thebibliography}

\end{document}